\documentclass[12pt,a4paper]{article}
\pdfoutput=1
\usepackage{cite}
\usepackage{amsmath}
\usepackage{color}
\usepackage{amsfonts}
\usepackage{amssymb}
\usepackage{graphicx}
\usepackage{geometry}
\usepackage{amssymb,epsfig,subfigure}
\usepackage{hyperref}
\usepackage{comment}
\usepackage[font=footnotesize]{caption}

\makeatletter
\renewcommand\section{\@startsection {section}{1}{\z@}%
                                 {-3.5ex \@plus -1ex \@minus -.2ex}
                                   {2.3ex \@plus.2ex}%
                                   {\normalfont\large\bfseries}}
\renewcommand\subsection{\@startsection{subsection}{2}{\z@}%
                                   {-3.25ex\@plus -1ex \@minus -.2ex}%
                                     {1.5ex \@plus .2ex}%
                                     {\normalfont\bfseries}}
\renewcommand\subsubsection{\@startsection{subsubsection}{3}{\z@}%
                                   {-3.25ex\@plus -1ex \@minus -.2ex}%
                                     {1.5ex \@plus .2ex}%
                                     {\normalfont\itshape}}
\makeatother

\def\pplogo{\vbox{\kern-\headheight\kern -29pt
\halign{##&##\hfil\cr&{\ppnumber}\cr\rule{0pt}{2.5ex}&\ppdate\cr}}}
\makeatletter
\def\ps@firstpage{\ps@empty \def\@oddhead{\hss\pplogo}%
  \let\@evenhead\@oddhead 
}
\def\maketitle{\par
 \begingroup
 \def\thefootnote{\fnsymbol{footnote}}
 \def\@makefnmark{\hbox{$^{\@thefnmark}$\hss}}
 \if@twocolumn
 \twocolumn[\@maketitle]
 \else \newpage
 \global\@topnum\z@ \@maketitle \fi\thispagestyle{firstpage}\@thanks
 \endgroup
 \setcounter{footnote}{0}
 \let\maketitle\relax
 \let\@maketitle\relax
 \gdef\@thanks{}\gdef\@author{}\gdef\@title{}\let\thanks\relax}
\makeatother

\numberwithin{equation}{section}

\newcommand\eea{\end{eqnarray}}
\newcommand\bea{\begin{eqnarray}}

\def\beq{\begin{equation}}
\def\eeq{\end{equation}}

\newcommand{\be}{\begin{equation}}
\newcommand{\ee}{\end{equation}}
\newcommand{\ba}{\begin{align}}
\newcommand{\ea}{\end{align}}
\newcommand{\bg}{\begin{gather}}
\newcommand{\eg}{\end{gather}}
\newcommand{\bseq}{\begin{subequations}}
\newcommand{\eseq}{\end{subequations}}

\renewcommand{\t}{\tilde}

\newcommand{\mc}{\mathcal}

\begin{document}
\setcounter{page}0
\def\ppnumber{\vbox{\baselineskip14pt
}}
\def\ppdate{
} \date{}

\author{Horacio Casini, Eduardo Test\'e, Gonzalo Torroba\\
[7mm] \\
{\normalsize \it Centro At\'omico Bariloche and CONICET}\\
{\normalsize \it S.C. de Bariloche, R\'io Negro, R8402AGP, Argentina}
}

\bigskip
\title{\bf  All the entropies on the light-cone 
\vskip 0.5cm}
\maketitle

\begin{abstract}
We determine the explicit universal form of the entanglement and Renyi entropies, for regions with arbitrary boundary on a null plane or the light-cone. All the entropies are shown to saturate the strong subadditive inequality. This Renyi Markov property implies that the vacuum behaves like a product state. For the null plane, our analysis applies to general quantum field theories, and we show that the entropies do not depend on the region. For the light-cone, our approach is restricted to conformal field theories. In this case, the construction of the entropies is related to dilaton effective actions in two less dimensions. In particular, the universal logarithmic term in the entanglement entropy arises from a Wess-Zumino anomaly action. We also consider these properties in theories with holographic duals, for which we construct the minimal area surfaces for arbitrary shapes on the light-cone. We recover the Markov property and the universal form of the entropy, and argue that these properties continue to hold upon including stringy and quantum corrections. We end with some remarks on the recently proved entropic $a$-theorem in four spacetime dimensions.
\end{abstract}
\bigskip

\newpage

\tableofcontents

\vskip 1cm

\section{Introduction}\label{sec:intro}

Quantum information theory provides powerful techniques to understand nonperturbative aspects of quantum field theory (QFT). One useful way in which this has worked out is by applying information-theoretic inequalities, such as strong subadditivity or monotonicity of the relative entropy, to QFT. These inequalities give insights into causality and unitarity constraints in relativistic theories, which are often hard to recognize from local observables. Some examples include energy conditions in QFT~\cite{Blanco:2013lea, Faulkner:2016mzt, Balakrishnan:2017bjg, Bousso:2015mna, Bousso:2015wca, Koeller:2015qmn, Leichenauer:2018obf}, and proofs of the irreversibility of renormalization group (RG) flows in various dimensions~\cite{Casini:2004bw, Casini:2012ei,  Casini:2016fgb, Casini:2016udt, Casini:2017vbe, Lashkari:2017rcl}.

\begin{figure}[h!]
\begin{center}  
\includegraphics[width=0.6\textwidth]{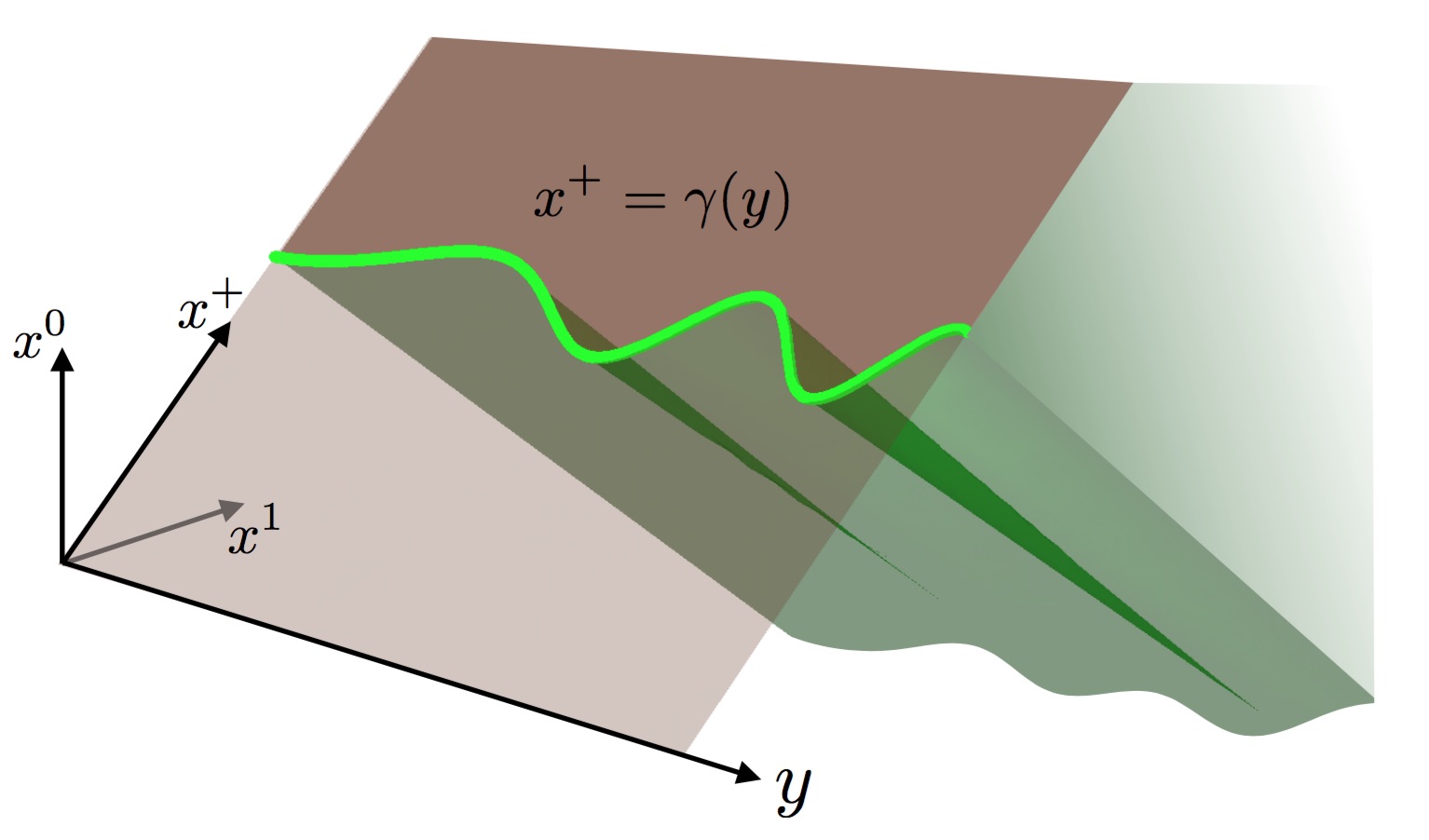}
\captionsetup{width=0.9\textwidth}
\caption{Region with boundary $x^+=\gamma(y)$ (green curve) on the null plane  $x^-=0$ and parallel to $k=(1, 1, 0, \ldots)$. Here $y$ are the $d-2$ transverse coordinates.
}
\label{fig:plane}
\end{center}  
\end{figure}

\begin{figure}[t]
\begin{center}
\includegraphics[width=0.45\textwidth]{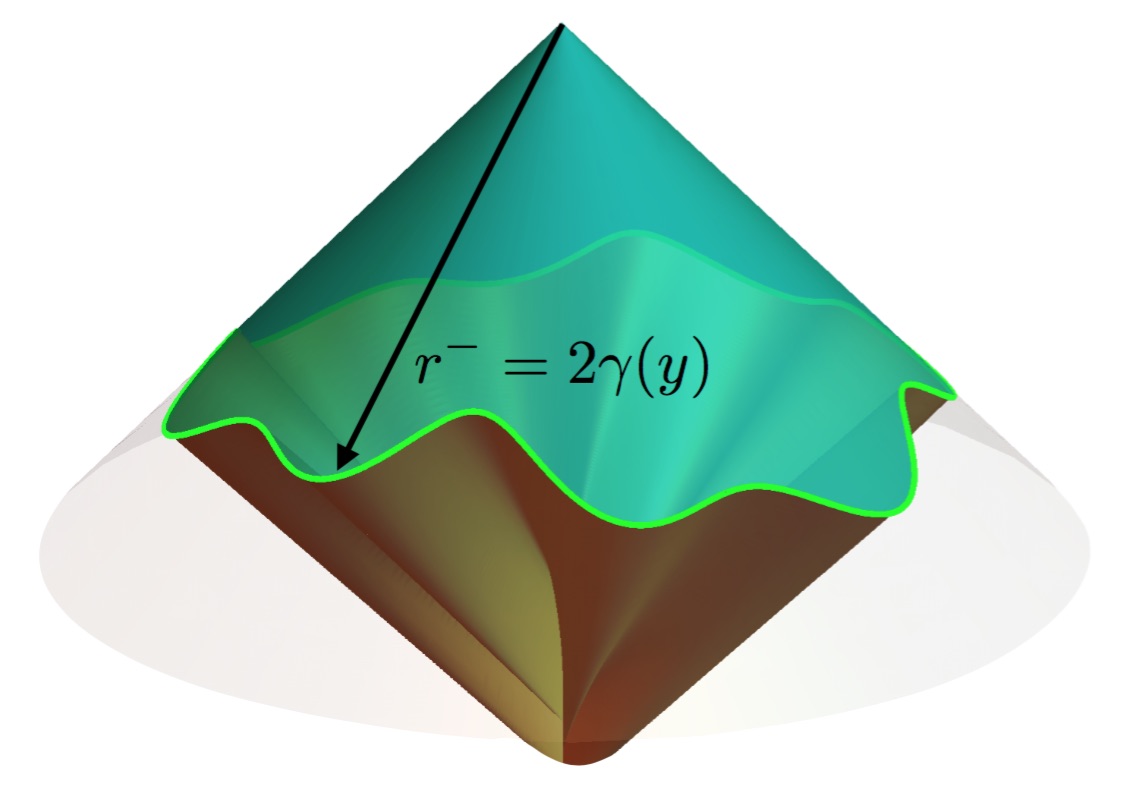} 
\caption{A region with boundary on the light-cone. This setup applies for CFTs.}
\label{fig:cone}
\end{center}
\end{figure}

Recently, it has become clear that these results can be extended and generalized by taking the null limit.\footnote{This was motivated by the entropic proof of the $g$-theorem in~\cite{Casini:2016fgb}, which recognized that working with Cauchy surfaces that approach the null cone allows to derive nontrivial constraints for the irreversibility of the RG. See also~\cite{Casini:2016udt}.} Here one considers the reduced density matrix $\rho_X$ for a region $X$ whose boundary $\gamma$ lies on a null plane or on the light-cone. See Figs.~\ref{fig:plane} and \ref{fig:cone}. For these regions, Ref.~\cite{Casini:2016udt} obtained the modular Hamiltonian, which turns out to be local and given by the Rindler result, ray by ray. See also \cite{Koeller:2017njr,Wall:2011hj}. This surprising result is a consequence of the special geometry and symmetries on the null plane. As a consequence, the entanglement entropy (EE) for general QFTs saturates the strong subadditive (SSA) inequality on the null plane,
\be\label{eq:Markov}
S_A+S_B-S_{A\cap B}-S_{A\cup B}=0\,. 
\ee
This is called the Markov property, in analogy with the classical case. For a conformal field theory (CFT), the null plane can be mapped to the light-cone, and then (\ref{eq:Markov}) holds on the null cone as well. With this result for CFTs, we showed in~\cite{Casini:2017vbe} that for RG flows between UV and IR fixed points, the change $\Delta S(r) = S(r) - S_{CFT_{UV}}(r)$ in the EE for a sphere obeys
\be\label{eq:beautiful}
r\, \Delta S''(r) -(d-3) \Delta S'(r)\le 0\,.
\ee 
This leads to a new proof of the $a$-theorem in four spacetime dimensions, and it also reproduces the proof of~\cite{Casini:2012ei} for the $c$-theorem in two dimensions and the $F$-theorem in three dimensions. In this way, a single formula unifies all known results for the irreversibility of the RG in Lorentz invariant QFTs in $d \le 4$. See also~\cite{Lashkari:2017rcl} for related work.

In the present work, we will analyze in detail the explicit form of the entanglement and Renyi entropies for regions with arbitrary boundaries $\gamma$ on the null plane (for general QFTs) and on the light-cone (for CFTs). In Sec.~\ref{sec:markov} we will provide simple geometric arguments that will prove that the EE and all Renyi entropies are in fact \textit{independent} of $\gamma$ on the null plane. This is a very strong result, and it implies that all Renyi entropies also satisfy the Markov property (\ref{eq:Markov}). This infinite set of equations for the reduced density matrix basically says that the vacuum state behaves like a product state over the null plane. In this sense, the result is opposite in spirit to the Reeh-Schlieder theorem, that forbids such products over spatial regions. 

The situation is much richer for regions with boundary on the light-cone, and we study this in Sec.~\ref{sec:qft}. Using Lorentz invariance and the Markov property, we determine the universal explicit form for all the entropies as a function of $\gamma$. This generalizes the result for the EE of a sphere to arbitrary boundaries. We obtain a local functional that is an integral over the angular coordinates of the light-cone. We interpret this as an effective action for a dilaton $\log \gamma(y)$ in $d-2$ dimensions.\footnote{For earlier work connecting the EE to a dilaton field theory in two less dimensions see~\cite{Solodukhin:2013yha}.} In particular, we argue that the universal logarithmic term for the sphere EE generalizes to the Wess-Zumino anomaly action for the dilaton.

In the second part of the paper (Sec.~\ref{sec:holo}) we study these questions from the point of view of AdS/CFT.\footnote{The results of this section were presented by the authors during 2017 at various seminars and conferences.} The EE for the boundary theory becomes the area of the extremal Ryu-Takayanagi surface in the gravitational theory. We construct the extremal surfaces corresponding to regions of the boundary QFT on the null plane and the light-cone. This geometric problem turns out to have various special features: the surfaces are described by linear differential equations (bulk laplacians), and they lie themselves on the bulk null plane or cone. We verify that the Markov property holds holographically. For the null cone, we evaluate the holographic EE explicitly, and check that it agrees with a special case of the general form predicted for CFTs in Sec.~\ref{sec:qft}. These results are extended to include $1/N$ and 't Hooft coupling corrections.

Armed with these additional insights, in Sec.~\ref{sec:athm} we revisit the proof of the $a$-theorem of~\cite{Casini:2012ei}, checking and expanding on the arguments in that work. In the process, we uncover a new positivity constraint for a nonlocal term in the EE. Lastly, in Sec.~\ref{sec:concl} we discuss implications of our results and various future directions.

\textit{Note added:} while we were preparing the manuscript for submission, the work~\cite{Neuenfeld:2018dim} appeared, which also studies extremal surfaces with boundaries on the null plane and cone in holographic theories. Some of the results in Sec.~\ref{sec:holo} -- specifically, our formulas (\ref{con}) and (\ref{eq:solrt}) -- overlap with that reference.

\section{Markov property for Renyi entropies}
\label{sec:markov}

In~\cite{Casini:2017roe} we showed that modular Hamiltonians $H_X$ for regions $X$ with boundary on a null plane $x^-=x^1-x^0=0$ are given by
\be
H_\gamma=2\pi \int d^{d-2} y\, \int_{\gamma(y)}^\infty dx^+\, (x^+-\gamma(y)) T_{++}(\lambda, y)\,, \label{eq:DeltaH2}
\ee
up to an additive constant. Here $y$ denote the transverse coordinates $(x^2, \ldots, x^{d-1})$, and $x^+= \gamma(y)$ parametrizes the boundary of $X$ on the null plane. This is simply the Rindler result, ray by ray. It leads to the operator equation
\be
H_A+H_B-H_{A\cap B}-H_{A\cup B}=0\,,
\ee
which in turn implies the Markov property for the entanglement entropies (EE) 
\be
S_A+S_B-S_{A\cap B}-S_{A\cup B}=0\,. \label{mcom}
\ee

In this section we will prove a much stronger statement, namely that all vacuum Renyi entropies of regions with boundary on the null plane also satisfy the Markov property. Our analysis on the null plane will be valid for any QFT. Hence, for conformal field theories (CFT), after a conformal transformation, the Markov property also holds for Renyi entropies of regions with boundary on the null cone. This gives an infinite set of equations for the vacuum reduced density matrix, placing strong constraints on quantum entanglement in QFTs.

We will argue that these properties for the entropies arise simply from geometrical considerations. In fact, our arguments also extend to other quantities such as free energies with insertions of $(d-2)$ dimensional surface operators. In the future, it would be interesting to understand the implications of our formulas for surface operators in gauge theories.

\subsection{Proof of the Markov property}\label{subsec:proof}

Let us first describe the setup in more detail. We work in $d$-dimensional Minkowski space with signature $(-,+,\ldots, +)$, and introduce null coordinates
\be
x^\pm = x^1 \pm x^0\,.
\ee
Consider a null plane $x^-=0$ with orthogonal coordinates $x^+$ and $y^a=(x^2,\ldots,x^{d-1}) \in \mathbb R^{d-2}$. The metric on the plane is
\be\label{eq:null-metric}
ds^2=(dy^a)^2+0 \, dx^+ dx^-\,.
\ee
We take a $d-2$ dimensional surface $x^+=\gamma(y)$ on the null plane, crossing all null rays --see Fig.~\ref{fig:plane}.

We wish to compute the vacuum entanglement Renyi entropy $S_n$ of a QFT in a region with boundary in $\gamma(y)$. Since the entanglement entropy does not depend on the Cauchy surface but on the whole causal region, it is equivalent to say that it is a functional of the boundary $\gamma(y)$. 
We assume a Lorentz invariant regularization of the entropies, with short distance cutoff $\epsilon$. A Lorentz invariant cutoff can be produced using the mutual information, or mutual Renyi entropies; see Appendix \ref{Appendix:mutual}.   
In a theory with mass scales, $S_n$ can also depend on other dimensionful parameters. Since we are working with the vacuum state, we can only use the geometry of $\gamma$, $\epsilon$, and some constants of the theory to construct $S_n(\gamma)$. In particular, we can expand in terms of functionals of the form
\be\label{eq:functional}
S_n(\gamma)=\int d^{d-2}\sigma_{y_1}\,\ldots \int  d^{d-2}\sigma_{y_n} \,f(\gamma(y_1),\ldots,\gamma(y_n); \nabla \gamma(y_1), \ldots)\,, 
\ee  
where $d\sigma$ is a volume element along $\gamma$ and $f$ is a function of the distances between points and the dimensionful parameters. 

The simplest argument is as follows.
These functionals should be Lorentz invariant. In particular, a boost rescales the coordinate $x^+ \to \lambda x^+$, so we have
\be
S_n(\gamma)=S_n(\lambda \gamma)\,, 
\ee   
for any $\lambda>0$. Taking the limit $\lambda\rightarrow 0$, and focusing on bounded curves, the entropy of $\gamma$ must then be the same as the one of a surface arbitrarily near the plane $x^+=0$.\footnote{We are implicitly neglecting some ``pathological" Lorentz invariant functionals which still distinguish smooth surfaces arbitrarily close (along with all the derivatives) to $\gamma=0$, as the one counting the number of maximums in $\gamma$. We expect the cutoff entropies should be continuous as functions of the shape in this sense.} Therefore, $S_n$ must be independent of $\gamma$. 

Another way to establish this is to realize that the degenerate metric (\ref{eq:null-metric}) gives an infinite set of isometries for the null plane
\bea
y &=& y'\,, \nonumber \\
x^+ &=& h(y',  {x^+}')\,. \label{defo}
\eea
That is, we can deform the $x^+$ coordinate in a way dependent on $y$, and get the same metric. These are of course not isometries of the full Minkowski space. Any two surfaces $\gamma$ can be deformed into one another by these isometries. Hence they have identical (flat) intrinsic geometry and also they are identically embedded in the null plane. These isometries imply that the functional (\ref{eq:functional}) will be the same for all $\gamma$. Nothing changes if we consider using derivatives of $\gamma$ of any order to form the functional of $\gamma$.  More explicitly, multiple gradients of $\gamma$ are tensors that can be expanded with the orthogonal vectors $k=(1,1,0,\ldots, 0)$ and $\hat{y}^a$, and the same holds for the distance vectors between any two points along $\gamma$. Once these tensors are contracted the components proportional to $k$ do not contribute because $k^2=0$, $k \cdot \hat{y}^a=0$. Hence the remaining contribution is the same as the one of a planar $\gamma$, and hence independent of the shape of $\gamma$. 

Another aspect of this impossibility of distinguishing different $\gamma$ with a geometric functional is that we cannot form non trivial invariants from the extrinsic curvatures of $\gamma$. There are two null vectors normal to $\gamma$, $k=(1,1,0,\ldots,0)$ and $q$, $q^2=0$, normalized with $k\cdot q=1$. Since $k$ is constant along $\gamma$, the corresponding extrinsic curvature vanishes. There is an ambiguity $k\rightarrow \lambda k$, $q\rightarrow 1/\lambda q$ in the representation of the surface in terms of the orthogonal null vectors. Then, in order to produce an invariant we have to use products of curvatures for $q$ and $k$, which are also zero.  

We conclude that all functionals we can construct should give the same value of $S_n$ for any $\gamma$.\footnote{For the entropy, this statement might be related, in an admittedly obscure way, with a similar statement in \cite{Casini:2017roe} for infinite dimensional systems where the Markov property holds for the full modular Hamiltonians.}  
The Markov property for $S_n$ then follows trivially, that is, the combination   
\be\label{eq:Renyi-Markov}
S_n(A)+S_n(B)-S_n(A\cap B)-S_n(A\cup B)=0\,,
\ee
because all the entropies are equal. 

This result for the independence of $S_n$ on $\gamma$ did not assume any unitary symmetry of the vacuum corresponding to the deformations (\ref{defo})  of the null plane.
 However, in addition to Lorentz boosts, such unitary symmetries deforming the null plane along the null rays and keeping the vacuum invariant do indeed exist for the special case $x^+=x^{+ \, \prime}+ \gamma^\prime(y^\prime)$. These are given by the modular translations corresponding to other arbitrary regions $\gamma^\prime$ with boundary in the null plane \cite{Casini:2017roe}. They act as isometries on the plane but do not have local action on field operators outside the plane. Therefore, the transformations between different surfaces $\gamma$ can indeed be implemented by unitaries keeping the vacuum invariant.  

This geometric argument implies that the equality of the entropies for all $\gamma$ extends to other quantities such as partition functions with insertions of $d-2$ dimensional surface operators. But this does not apply to lower dimensional operators which are not equivalent under the isometries of the null plane.  

The argument above needed a Lorentz invariant cutoff. Once this requirement is dropped the equality of all entropies for different $\gamma$ does not hold any more -- we could for example change the cutoff around $\gamma$ and $\gamma^\prime$ independently. However, the Markov property is a regularization independent statement. The reason is that the divergences in the entropies are local and extensive on the boundary of the region; hence in any other regularization they must also cancel locally in the combination (\ref{eq:Renyi-Markov}).

In conclusion, a Lorentz invariant geometric functional of $d-2$ surfaces with minimal continuity properties must be constant on regions with boundary on a null plane. If this functional is either finite or has local extensive divergences along $\gamma$, it must be Markovian on the null plane, and this is a cutoff independent statement. This property then persists on the null cone for a conformally invariant functional (that is, a functional that is conformally invariant for any cutoff independent combination). 

We will next illustrate this with a model having extensive mutual information. We will also see directly this structure for the holographic entanglement entropy in Sec.~\ref{sec:holo}. 

\subsection{An example: extensive mutual information model}\label{subsec:EMI}

A simple example is given by the EMI (extensive mutual information) model for the entropy~\cite{Casini:2008wt}. For a spatial surface $A$ with complement $\bar{A}$ in a given Cauchy surface, this model gives the functional 
\be
S(A)=\int_A d\sigma_x\, \int_{\bar{A}} d\sigma_y \, \eta_x^\mu \,\eta_y^\nu \,(\partial_\mu \partial_\nu-g_{\mu\nu} \partial^2) \,|x-y|^{-(2d-4)}\,, \label{aa}
\ee
where $\eta$ is the normalized vector orthogonal to the Cauchy surface. A small distance cutoff is assumed between $A$ and $\bar{A}$. The interest of this expression is that it gives a simple example of conformal invariant, positive, and strong subadditive functional on causal regions. It can also be thought of as the free energy in the presence of surface operators which are exponentials of free fields~\cite{Swingle:2010jz}.

The integrand is a conserved current in both indices what guarantees $S$ is independent of the Cauchy surface. In fact this expression is equivalent to one dependent only on the boundary of $A$
\be
S(A)=\int_{\partial A} d\sigma_x^{\alpha\beta}\, \int_{\partial A} d\sigma_y^{\alpha\beta}\, \frac{1}{|x-y|^{2(d-2)}}\,,\label{sis}
\ee
where again a small cutoff is assumed at coincidence points.
With a distance cutoff in (\ref{sis}), a quick look at the argument above confirms  $S$ is independent of the region on the null plane. Markovianity on the cone can be seen directly from  (\ref{aa}), choosing the null cone as a Cauchy surface. Then the Markov combination (\ref{mcom}) reduces to the (finite) double integral of the integrand in (\ref{aa}) over non-overlapping regions $A\cap \bar{B}$ and $B\cap \bar{A}$ of the null cone. It is easy to check explicitly that the double integral over patches of the same null cone vanishes identically, while it is always positive for other null patches or spatial regions. This vanishing gives the Markovian property for this functional.

\section{Universal form of CFT entropies on the light-cone}
\label{sec:qft}

In this section we study the vacuum reduced density matrix for regions whose boundary lies on the light-cone. We
will determine the universal form of the entanglement and Renyi entropies for general CFTs.

The conformal transformation between the plane and the cone, working in the metric with signature $(-+...+)$, is given by 
\be\label{maa}
x^\mu=2\frac{X^\mu+(X\cdot X)C^\mu}{1+2(X\cdot C)+(X\cdot X)(C\cdot C)}-D^\mu\,,\ \ \ \ C^\mu\equiv(0,1/R,\vec{0})\,,\ \ \ \ D^\mu=(R,R,\vec{0}) .
\ee 
This maps the past light-cone of the origin $x^\mu=0$ into (part of) the null plane $X^-=X^1- X^0= 0$. The origin $X^\mu=0$ is mapped into the point $(-R,-R,\vec{0})$, the surface $X^\pm=0$ is mapped to the circle $x^0=-R$, $r=R$. The points on the null cone from the point line $x^1=-x^0=R$ correspond to the infinity in the coordinates $X$. We will then consider a surface\footnote{To simplify notation, the boundaries on the null plane and cone are denoted as $\gamma$.}
\be
r^- = 2 \gamma(y)
\ee
on the past light-cone $r^+=0$, with 
\be
r^\pm = r \pm x^0\,.
\ee 
This curve parametrizes the boundary of the Cauchy surface. The restriction of the Minkowski metric to $r^+=0, r^-=2 \gamma(y)$ gives a $(d-2)$-dimensional sphere with radius that depends on the angular position along the curve:
\be\label{eq:ds0}
ds^2= 0\,dr^+ dr^- + \gamma(y)^2\,g_{ab}(y) dy^\alpha dy^\beta\,.
\ee
Here 
\be\label{eq:sphere-metric}
g_{ab}(y) dy^a dy^b= \frac{4}{(1+y^2)^2}\,(dy^a)^2
\ee
describes a sphere $S^{d-2}$ of unit radius in conformally flat coordinates.\footnote{To see this, change variables to $y^a= \tan(\theta/2)\, \hat n^a$, with $\hat n^a$ unit vectors.}

We argued in the previous section that the entropies for a Cauchy surface with boundary on the null plane and Lorentz invariant regularization are independent of the boundary shape. After a conformal transformation to the light-cone, this means that all the dependence on $\gamma$ has to arise from the short-distance cutoff $\epsilon$ on the light-cone. (We will see explicit examples of this in holographic theories in Sec.~\ref{sec:holo}). Up to an overall constant, this is local and extensive, and hence the entanglement and Renyi entropies should be given by local functionals of $\gamma/\epsilon$,  its derivatives, and geometric quantities built from $g_{ab}$
\be\label{eq:localS}
S_n= \int d^{d-2}y\,\sqrt{g}\,L_n(\gamma/\epsilon, g_{ab}, \partial \ldots)+F_n\,.
\ee
Equivalently, the Markov property on the null plane is regularization invariant and hence preserved by the conformal transformations for a CFT. The Markov property on the null cone implies that the entropy is a local functional plus possibly a constant $F_n$ independent of $\gamma$. 

Our goal is to determine the general form of $L_n$ allowed by Lorentz invariance. We will find that this is related to a dilaton effective action on $S^{d-2}$. Our analysis will reveal how the EE for spheres
\be\label{eq:EEsphere}
S(\gamma)=\alpha_{d-2}\,\frac{\gamma^{d-2}}{\epsilon^{d-2}}+\alpha_{d-4}\, \frac{\gamma^{d-4}}{\epsilon^{d-4}}+\ldots+ \left\lbrace \begin{array}{ll} (-)^{\frac{d}{2}-1} 4\,A\, \log(\gamma/\epsilon)\,& d\;  \textrm{even}\,.\\ (-)^{\frac{d-1}{2}} F\,& d\,\,\textrm{odd} \,. \end{array}\right.
\ee   
generalizes to an arbitrary boundary $\gamma(y)$ on the light-cone. The main results are given in (\ref{eq:Sodd}) and (\ref{eq:Seven}). The divergent terms are automatically Markovian, and we will find the form of the universal finite contributions.

\subsection{Lorentz transformations on the light-cone}\label{subsec:lorentz}

In order to impose Lorentz invariance, we need to determine how Lorentz transformations act on the subspace $r^+=0,\,r^-= 2\gamma(y)$. The pull-back metric is (\ref{eq:ds0}), which describes an $S^{d-2}$ with varying radius $\gamma(y)$. It is known that Lorentz transformations reduce to conformal transformations on $S^{d-2}$; this becomes clear in the embedding space formalism, where conformal transformations are represented as linear transformations on a null-cone of a projective space in two more dimensions. We will now review how this comes about; see e.g.~\cite{Weinberg:2010fx, Penedones:2016voo}.

It is useful to parametrize the null cone $\mathcal C$ as
\be
x^\mu(\lambda, y^a)= \lambda\, \omega(y)\,\hat x^{\mu}(y)\;,\;\hat x^\mu(y) = \left(\frac{1+ y^2}{2}, y^a,\frac{1- y^2}{2} \right)\,,
\ee
where $\lambda \in \mathbb R$, $y^a \in \mathbb R^{d-2}$. The coordinate $\hat x^\mu$ gives the Poincar\'e section $\hat x^0+ \hat x^d=1$ of the null cone $\eta_{\mu\nu} \hat x^\mu \hat x^\nu=0$; $\lambda$ describes `radial' motion on the cone. See also~\cite{Kapec:2017gsg}. The conformal factor $\omega(y)$ can be arbitrary but here we will fix it to
\be
\omega(y)=\frac{2}{1+y^2}\,.
\ee
The pull-back of the Minkowski metric to $\mathcal C$ then reads
\be\label{eq:dsy}
ds^2_{\mathcal C}= \lambda^2 \frac{4}{(1+y^2)^2} (dy^a)^2\,,
\ee
which, recalling (\ref{eq:sphere-metric}), describes a sphere in conformally flat coordinates. In particular, we are interested in a sphere of varying radius $\gamma(y)$, and this is obtained for
\be\label{eq:sphere-coords}
\lambda = \gamma(y)\,.
\ee

The main advantage of these coordinates is that there is a simple relation between Lorentz transformations on $x^\mu$ and conformal transformations on $(\lambda, y^a)$. In more detail, the Lorentz generators $J_{\mu\nu}$ induce $SO(d-2)$ rotations, translations, special conformal transformations and dilatations on $\mathcal C$:
\be
J_{ab}\;,\;T_a= J_{0,a}- J_{d-1,a}\;,\;K_a=J_{0,a}+ J_{d-1,a}\;,\;D= J_{d-1,0}\,.
\ee
In this way, the Lorentz algebra $SO(d-1,1)$ gives rise to the conformal algebra for euclidean $\mathbb R^{d-2}$. The coordinates transform as $(\lambda, y) \to (\lambda', y')$ with
\be\label{eq:conf}
\frac{\partial y'^a}{\partial y^c}\frac{\partial y'^b}{\partial y^d} \delta_{ab} =  e^{2A(y)}\delta_{cd}\;,\;\lambda'= e^{-A(y)} \lambda\,.
\ee
Note that while the embedding space $\mathbb R^{d-1,1}$ for CFTs is just an artifact, in our setup it is the physical space where the QFT lives.

\subsection{Entropies on the null cone}\label{subsec:univ}

Our goal now is to determine the general form of (\ref{eq:localS}) consistent with Lorentz invariance.  We can think of $S_n$ as an ``action'' for an euclidean theory that lives on $S^{d-2}$, with a scalar degree of freedom $\gamma(y)$. As reviewed in Sec.~\ref{subsec:lorentz}, Lorentz transformations act as conformal transformations on $S^{d-2}$, so we will keep the metric $g_{ab}$ explicit to account for conformal rescalings, which act as $g_{ab} \to e^{2A(y)} g_{ab}$. Furthermore, from (\ref{eq:conf}), $\phi(y)=\log (\gamma(y)/\epsilon)$ transforms additively as a dilaton field.
In this way, the problem of finding the entropies $S_n$ is equivalent to that of constructing a conformally-invariant local action in $d-2$ dimensions with a dilaton field $\phi(y)=\log (\gamma(y)/\epsilon)$. 

It is interesting to note that dilaton techniques have appeared in the recent proof of the $a$-theorem in~\cite{Komargodski:2011vj}; see also~\cite{Schwimmer:2010za, Komargodski:2011xv, Elvang:2012st, Luty:2012ww}. There, the dilaton is introduced by hand in order to match Weyl anomalies; in our context $\phi(y)$ is physical, as it arises from the varying radius of $S^{d-2}$ on the light-cone. These results on the dilaton effective action will be useful for our goal, especially the $d$-dimensional analysis in~\cite{Elvang:2012yc}.\footnote{Dilaton methods have also been used in EE calculations in~\cite{Banerjee:2011mg, Solodukhin:2013yha, Banerjee:2014daa, Herzog:2015ioa}.}

\subsubsection{Odd $d$}

Let us begin with the simpler case of odd space-time dimension $d$. The `action' functional for the entropy $S_n(\gamma)$ can be constructed simply as a derivative expansion in terms of local geometric invariants built from the metric
\be\label{eq:hatg}
\hat g_{a b} \equiv \frac{\gamma(y)^2}{\epsilon^2}\,g_{ab}(y)\,,
\ee
with $g_{ab}$ the metric of the unit radius $S^{d-2}$. Since this is the metric induced by the Minkowski metric on $\gamma$ it is clear that these geometric terms are Lorentz invariant.
We note that the Riemann tensor can be written in terms of $\hat R_{ab}$ and $\hat R$ because $\hat g_{ab}$ is conformally flat (the Weyl tensor vanishes). In addition we could construct invariants using the extrinsic curvatures of $\gamma$. We show in Appendix \ref{Appendix:uno} that the extrinsic curvatures on the null cone give again combinations of the intrinsic metric and the Ricci tensors.

Thus the most general effective action is constructed in terms of powers of $\hat g_{ab}$, the Ricci tensor, the Ricci scalar and covariant derivatives.
The first few terms are
\be\label{eq:Snodd}
S_n(\gamma) = \int d^{d_\perp} y\,\sqrt{\hat g}\left(\beta_0 + \beta_2 \hat R+ \beta_4 \hat R^2 + \beta_4'\, (\hat R_{\alpha \beta})^2+ \ldots \right)+F_n\,,
\ee
with $d_\perp \equiv d-2$. The constant coefficients $\beta_j$ depend on the specific theory and on $n$. In this expression, conformal invariance for the dilaton --namely Lorentz invariance for the $d$-dimensional QFT-- is manifest. 

To gain intuition, let us write explicitly the terms with zero and two derivatives:
\bea
\int d^{d_\perp}y\,\sqrt{\hat g} &=& \int d^{d_\perp}y\,\sqrt{g}\,\frac{\gamma(y)^{d_\perp}}{\epsilon^{d_\perp}} \,,\\
\int d^{d_\perp}y\,\sqrt{\hat g} \,\hat R &=& \int d^{d_\perp}y\,\sqrt{g}\,\frac{\gamma^{d_\perp-2}}{\epsilon^{d_\perp-2}} \left((d_\perp-1)(d_\perp-2) \left( \frac{\nabla \gamma}{\gamma}\right)^2+ d_\perp (d_\perp-1) \right)\,. 
\eea
The first term is the familiar area term. Performing a field redefinition
\be
\varphi(y) = 2 \sqrt{\frac{d_\perp-1}{d_\perp-2}}\,\left(\frac{\gamma(y)}{\epsilon} \right)^{(d_\perp-2)/2}\,,
\ee
the second term becomes,  for $d \ge 5$, the action for a conformally coupled scalar,
\be\label{eq:conformalkin}
\int d^{d_\perp}y\,\sqrt{\hat g} \,\hat R=\int d^{d_\perp}y\,\sqrt{g}\, \left((\nabla \varphi)^2+ \xi R \varphi^2 \right)\,,
\ee
where $\xi = \frac{d_\perp-2}{4(d_\perp-1)}$ and the Ricci scalar $R=d_\perp(d_\perp-1)$ for the unit-radius sphere.\footnote{On the other hand, this term vanishes for $d=2,3$ and is proportional to the volume of $S^{d-2}$ in $d=4$.} The area term proportional to $\gamma^{d_\perp}$ is then simply a conformal potential $V(\varphi) \sim \varphi^{2d_\perp/(d_\perp-2)}$. The next terms in the `effective action' for the entanglement entropy $S$ are higher derivative generalizations of this conformal Laplacian --we will return to this point below.

Note that the overall constant $F_n$ is trivially consistent with the Markov property (\ref{eq:Renyi-Markov}). However, it is not possible to write it as a local geometric invariant. In this sense it is analogous to the anomaly contributions for even $d$ to be discussed below. For entanglement over spheres, this is the familiar constant term $F$ that measures the free energy of the theory over the euclidean sphere.

Putting these results together, and replacing $d_\perp \to d-2$, the universal form of the EE for regions with boundary on the null cone and in odd space-time dimensions becomes
\bea\label{eq:Sodd}
S_n(\gamma)&=& \int\,d^{d-2}y\,\sqrt{g}\, \Bigg \lbrace \beta_0 \frac{\gamma(y)^{d-2}}{\epsilon^{d-2}} + \beta_2\frac{\gamma^{d-4}}{\epsilon^{d-4}} \left((d-2) (d-3)+(d-3)(d-4) \left(  \frac{\nabla \gamma}{\gamma}\right)^2 \right) \nonumber \\
&+& \ldots \Bigg \rbrace+ F_n\,.
\eea
Let us compare this with the EE for a CFT on a sphere, Eq.~(\ref{eq:EEsphere}).
We recognize in (\ref{eq:Sodd}) the area terms and all the subleading contributions, generalized to an arbitrary varying curve $\gamma(y)$. Some of the $\beta_k$ are fixed in terms of the entropy of the sphere. For instance, $\beta_0= \alpha_{d-2}, \beta_2=\alpha_{d-4}$. This means that the coefficient of $(\nabla \log \gamma)^2$ in the first subleading term $(\gamma/\epsilon)^{d-4}$ is uniquely fixed by the corresponding term in the sphere EE. This is a consequence of Lorentz invariance. At higher orders, there are more geometric invariants allowed, such as the terms with $\beta_4, \beta_4'$ in (\ref{eq:Snodd}). In this case, the sphere coefficient $\alpha_{d-2k}$ fixes only an overall combination of the $\beta_i$, and the entropy for the boundary $\gamma(y)$ contains more information about the specific theory. The term of order $\gamma^{d-2 - 2k}$ is essentially a higher-derivative version of the conformal Laplacian on the sphere containing $2k$ derivatives. We will discuss below a compact expression for such operators.

\subsubsection{Even $d$}

For $d$ even this is not the full story: there must be an additional contribution that comes from the Euler $a$-anomaly. Indeed, recall that for a sphere of constant radius $\gamma$ at fixed time, we should recover the universal logarithmic contribution
\be\label{eq:Sanom}
S_\text{anom} = (-1)^{d/2-1} 4 A \,\log \frac{\gamma}{\epsilon}\,.
\ee
We want to find a Lorentz invariant local functional that reduces to (\ref{eq:Sanom}) for constant $\gamma(y)$. At first, this appears to be challenging in our approach because, as we saw in (\ref{eq:Snodd}), there are no local invariants we can form with geometric quantities from $\hat g_{ab}$ that give rise to such a term.

We propose that the generalization of (\ref{eq:Sanom}) to arbitrary $\gamma(y)$ is a Wess-Zumino term for the Weyl anomaly on $S^{d-2}$. To explain how this comes about, let us first review the simplest case of the Weyl anomaly in 2d CFTs. The stress-tensor on a manifold with metric $g_{ab}$ has a trace-anomaly
\be
\langle T^a_a \rangle =\frac{c}{24\pi} R
\ee
where $R$ is the scalar curvature of $g_{ab}$. This implies that, under a Weyl rescaling $\delta g_{ab} =2 \delta \sigma g_{ab}$, the effective action $W = - \log Z$ changes as
\be\label{eq:deltaW}
\frac{\delta W}{\delta \sigma}  = - \frac{c}{24\pi}  R\,.
\ee
A local functional whose variation gives (\ref{eq:deltaW}) can be obtained by introducing a dilaton field $\tau$, which transforms as $\tau \to \tau+ \sigma(y)$ under $g_{ab} \to e^{2 \sigma(y)} g_{ab}$. The result is the Wess-Zumino action~\cite{Wess:1971yu}
\be\label{eq:SWZ}
S_\text{WZ}=  \frac{c}{24\pi}  \int d^2 y\,\sqrt{g}\left( \tau R -(\nabla \tau)^2\right)\,.
\ee
Here the dilaton derivative term cancels the Weyl transformation of the Ricci scalar, $R[e^{2\sigma} g]=e^{-2\sigma}(R[g]-2 \nabla^2 \sigma)$. We note that, while this is a local functional of $g_{ab}$ and $\tau$, it is not a local functional constructed from the Weyl-invariant metric $\hat g_{ab}= e^{-2 \tau} g_{ab}$.

Let us return now to the EE calculation for $d=4$.\footnote{We thank J. Maldacena for suggesting that the $d=4$ result can be mapped to a Liouville action.} We seek a local Lorentz-invariant functional that reduces to (\ref{eq:Sanom}) for constant $\gamma$. We found that Lorentz transformations act as conformal transformations on the $S^2$ null-cone sphere, and that $\log (\gamma/\epsilon)$ transforms as a dilaton field. We then recognize (\ref{eq:Sanom}) as the first term of the WZ action (\ref{eq:SWZ}) evaluated on $S^2$. In order to preserve Lorentz invariance, we expect that the contribution to the EE for a curve $\gamma(y)$ should then generalize to
\be\label{eq:SWZ4}
S_{WZ}=- \frac{A}{2\pi}\,\int d^2y \,\sqrt{g}\,\left(R\,\log\frac{\gamma(y)}{\epsilon}+ \left(\frac{\nabla \gamma}{\gamma}\right)^2 \right)\,,
\ee
with the overall normalization fixed by (\ref{eq:Sanom}) and the Euler characteristic $\frac{1}{4\pi}\int d^2 y \sqrt{g} R = 2$. Note that the coefficient of $\log(\epsilon)$ is topological and hence is the same for all $\gamma$. In particular, this means there is not type $B$ anomaly contribution to this logarithmic coefficient. This can be seen as a consequence of the particular geometry of the cone in Solodukhin's formula~\cite{Solodukhin:2008dh} for the coefficient of $\log(\epsilon)$ in generic regions in $d=4$. See the Appendix \ref{Appendix:uno}.

This is a local functional and hence satisfies the Markov property. But, as in the discussion of the Weyl anomaly, it is not a local functional of the metric $\hat g_{ab}= \frac{\gamma(y)^2}{\epsilon^2} g_{ab}$ introduced in (\ref{eq:hatg}). It is Lorentz invariant, as can be seen by writing it as a \textit{bilocal} functional~\cite{Deser:1993yx, Polchinski:1998rq}
\be\label{eq:Polch}
S_{WZ}\propto \int d^2 y \sqrt{\hat g}\,\int d^2 y' \sqrt{\hat g}\,\hat R(y) \hat{G}(y, y') \hat R(y')\,,
\ee
with $\nabla_y^2 \hat{G}(y,y') = \frac{1}{\sqrt{\hat g}}\,\delta^2(y,y')$ the Green's function for $\hat g_{ab}$, and $\hat R$ its curvature scalar. Using
\be
\sqrt{\hat g}\, \hat R = \sqrt{g} \left(R- 2 \nabla^2\log \frac{\gamma}{\epsilon} \right)
\ee
and integrating by parts, (\ref{eq:Polch}) reduces to (\ref{eq:SWZ4}), up to a term quadratic in $R$ that is independent of $\gamma$.

This discussion extends to arbitrary dimensions $d_\perp$, where the Weyl anomaly is proportional to the Euler density $E_{d_\perp}$ (plus conformally invariant terms that vanish in our case). The Wess-Zumino action can be computed systematically by integrating the Euler density~\cite{Wess:1971yu, Schwimmer:2010za},
\be\label{eq:SWZd}
S_{WZ}=(-1)^{d_\perp/2} \frac{4 A}{\chi_{d_\perp}}\,\int d^{d_\perp}y\,\sqrt{g}\,\int_0^1\,dt\,\log\frac{\gamma(y)}{\epsilon}\,E_{d_\perp}\left(\left(\frac{\gamma(y)}{\epsilon}\right)^{2t} g_{ab}\right)\,,
\ee
and $\chi_{d_\perp}= \int d^{d_\perp}y\,\sqrt{g}\,E_{d_\perp}(g)$ is proportional to the Euler character of the sphere. The contribution from $t=0$ reproduces (\ref{eq:Sanom}), and this is how the overall normalization is fixed. The full integral gives a conformally invariant action with derivatives of the schematic form $\int_y \log \frac{\gamma}{\epsilon} (\nabla^2)^{d_\perp/2} \log \frac{\gamma}{\epsilon}$. Explicit expressions in various even dimensions may be found in~\cite{Komargodski:2011vj, Komargodski:2011xv, Elvang:2012st, Elvang:2012yc, Herzog:2015ioa}. 

In summary, the entanglement entropy for an arbitrary curve $\gamma(y)$ in a CFT in even $d$ dimensions is given by
\bea\label{eq:Seven}
S_n(\gamma)&=& \int\,d^{d-2}y\,\sqrt{g}\, \Bigg \lbrace \beta_0 \frac{\gamma(y)^{d-2}}{\epsilon^{d-2}} + \beta_2\frac{\gamma^{d-4}}{\epsilon^{d-4}} \left((d-2) (d-3)+(d-3)(d-4) \left(  \frac{\nabla \gamma}{\gamma}\right)^2 \right) \nonumber \\
&+& \ldots + (-1)^{d/2-1} \frac{4 A_n}{\chi_{d-2}}\,\int d^{d-2}y\,\sqrt{g}\,\int_0^1\,dt\,\log\frac{\gamma(y)}{\epsilon}\,E_{d-2}\left(\left(\frac{\gamma(y)}{\epsilon}\right)^{2t} g_{ab}\right) \Bigg \rbrace\nonumber \\
&+& F_n\,.
\eea
The last term is the WZ action on $S^{d-2}$ with a dilaton $\log (\gamma/\epsilon)$, and it generalizes the universal logarithmic term of the EE on a sphere. In this case, $A_n = A$ is just the Euler anomaly.

For comparison with holographic results below, let us give some explicit examples. For $d=4$, using the curvature of $S^2$, $R=2$, we get, from (\ref{eq:SWZ4}),
\be\label{eq:SWZ4new}
S_{WZ}=- \frac{A}{2\pi}\,\int d^2\Omega\,\left(2\log\frac{\gamma(y)}{\epsilon}+ \left(\frac{\nabla \gamma}{\gamma}\right)^2 \right)\,,
\ee
Next, for $d=6$, we use that the WZ action (\ref{eq:SWZd}) becomes~\cite{Komargodski:2011vj}
\be
S_{WZ}=\frac{4 A}{\chi_{4}}\,\int d^{4}y\,\sqrt{g}\, \left(\phi E_4- 4(R_{ab}-\frac{1}{2} g_{ab} R)\partial_a \phi\, \partial_b \phi-4(\nabla \phi)^2 \nabla^2 \phi - 2 (\nabla \phi)^4\right)\,,
\ee
where $\phi = \log (\gamma/\epsilon)$.
Performing the calculation for a sphere obtains\footnote{Recall that for a maximally symmetric space in $n$ dimensions, $R_{\mu\nu}= \frac{R}{n} g_{\mu\nu}$, and $R_{\mu \nu\rho \sigma}= \frac{R}{n(n-1)}(g_{\mu\rho} g_{\nu \sigma}-g_{\mu \sigma} g_{\nu \rho})$~\cite{Weinberg:1972kfs}. Furthermore, for a unit-radius sphere $S^n$, $R= n(n-1)$. The Euler density in four dimensions reads $E_4= R_{\alpha \beta \mu\nu} R^{\alpha \beta \mu\nu}-4 R_{\mu\nu} R^{\mu\nu}+ R^2$, and for a sphere $E_4=24$. }
\be\label{eq:SWZ6}
S_{WZ}=\frac{3}{2\pi^2} A\,\int d^4 \Omega\,\left \lbrace \log \frac{\gamma}{\epsilon}+ \frac{1}{2}\left(\frac{\nabla \gamma}{\gamma} \right)^2 +\frac{1}{6}\left(\frac{\nabla \gamma}{\gamma} \right)^2 \left(\left(\frac{\nabla \gamma}{\gamma} \right)^2- \frac{\nabla^2 \gamma}{\gamma}\right)- \frac{1}{12}\left(\frac{\nabla \gamma}{\gamma} \right)^4 \right \rbrace\,.
\ee

\subsection{An alternative approach}\label{subsec:alternative}

We now present an alternative construction of the effective action. This approach is somewhat simpler, and makes it clear how Lorentz invariance of the $d$-dimensional theory is used.

First, we write the metric over the varying radius $S^{d-2}$ as a dilaton factor times the flat space metric,
\be
\frac{\gamma(y)^2}{\epsilon^2} d\Omega_{d-2}^2= e^{-2 \tau(y)} \,\delta_{ab} dy^a dy^b\;,\;e^{- \tau(y)} \equiv \frac{\gamma(y)}{\epsilon}\,\frac{2}{1+(y^a)^2}\,.
\ee
See discussion around (\ref{eq:sphere-coords}). We then require a local effective action, invariant under rotations and translations on $\mathbb R^{d-2}$, and under scale transformations $y \to e^{\sigma} y, \tau \to \tau+ \sigma$.

Following the construction of the dilaton effective action in~\cite{Elvang:2012yc}, this can be organized in terms of differential operators
\be\label{eq:Wk}
W_k = \left(\frac{2}{d_\perp-2k} \right)^2\,e^{- \frac{d_\perp-2k}{2} \tau} (\nabla^2)^k e^{- \frac{d_\perp-2k}{2} \tau}\,,
\ee
which contain $2k$ derivatives and transform covariantly under scale transformations,
\be
W_k \to e^{-d_\perp \sigma} W_k\,.
\ee 
Hence, the basic scale-invariant objects are $d^{d_\perp} y \,W_k$ and $e^{d_\perp \tau} W_r$, and the most general local effective action is
\be\label{eq:SeffW}
S_\gamma =\sum_{k, \bar{r}, \bar{n}} \int d^{d_\perp}y\,\alpha_{k \bar{r}}^{\bar{n}} \,W_k \,\prod_i \, (e^{d_\perp \tau} W_{r_i})^{n_i}\,,
 \ee
 with $\alpha_{k \bar r }^{\bar n}$ some arbitrary coefficients.
 The term proportional to $\alpha_{k \bar r}^{\bar n}$ contains $2k + 2 \sum_i n_i r_i$ derivatives.\footnote{We are including here all the terms allowed by scale invariance, while formula (2.43) in~\cite{Elvang:2012yc} contains only a subset of these terms. This is presumably because the effective action in that reference is evaluated on-shell for the dilaton, something which does not make sense in our context.}
 
 An explicit evaluation of the first few contributions in (\ref{eq:SeffW}) recovers the terms analyzed in Sec.~\ref{subsec:univ}. This approach has the advantage of unifying odd and even $d$; in particular, the Wess-Zumino term arises from the limit $k \to d_\perp/2$,
 \be
 \int\,d^{d_\perp} y \,W_{k=d_\perp/2} =\int\,d^{d_\perp} y \, \tau\,(\nabla^2)^{d_\perp/2} \tau\,.
 \ee
This is the reason for the normalization in (\ref{eq:Wk}).
 For instance, after integration by parts,
 \be
 \int d^2y\, \tau \,\nabla^2 \tau = \text{const}- \int d^2\Omega\,\left(2\log \frac{\gamma}{\epsilon}+\left(\frac{\nabla \gamma}{\gamma} \right)^2 \right)\,,
 \ee
 which agrees with (\ref{eq:SWZ4}).

\section{Holographic analysis}
\label{sec:holo}
 
In this section we analyze the entanglement entropy for regions with arbitrary boundaries on the null plane and, for CFTs, with arbitrary boundaries on the null cone, in theories with holographic duals. Via the HRT formula~\cite{Ryu:2006bv, Hubeny:2007xt}, this translates into finding extremal surfaces anchored at boundary curves $\gamma(y)$ in the null surfaces in asymptotically AdS space. This geometric problem turns out to have many special and interesting features, which are not present in the case of generic space-like boundary curves. In particular, we will find that the extremal surface is determined by a \text{linear} second order differential equation. We will check that the Markov property holds, and regain the general expressions of the previous section for EE in a null cone for CFTs. We will also show that these results hold when adding corrections for finite $N$ or finite 't Hooft coupling $\lambda$.

\subsection{Regions with boundary on a null plane}\label{subsec:plane}

The metric  for an asymptotically AdS space with Lorentz symmetry corresponding to the vacuum state in a holographic theory is
\be\label{eq:bulk-metric}
ds^2=\frac{L^2}{z^2}\left( f^2(z) dz^2 + dx^+ dx^- +d\vec{y}^2 \right)\,,
\ee
with $x^\pm=x^1\pm x^0$, $\vec{y}=(x^2,\ldots,x^{d-1})$, and $\lim_{z\rightarrow 0} f(z)=1$. Here $z\in (0,\infty)$ and $y^i\in (-\infty,\infty)$
We want to find an extremal surface in the bulk with boundary on a $d-2$ surface on the boundary given by
\be
x^-=0\;,\;x^+ =\gamma(\vec{y})\,.
\ee

The minimal surface has $d-1$ dimensions and we parametrize it with the coordinates $\alpha^i\equiv( z, \vec{y})$. The induced metric on this surface is
\be
h_{ij}=g_{\mu\nu}\frac{\partial x^\mu}{\partial \alpha^i}\frac{\partial x^\nu}{\partial \alpha^j} =\frac{L^2}{z^2} \left(\delta_i^1 \delta_j^1 (f^2(z)-1)+\delta_{ij}+\frac{1}{2} \left(\frac{\partial x^+}{\partial \alpha^i}\frac{\partial x^-}{\partial \alpha^j}+\frac{\partial x^-}{\partial \alpha^i}\frac{\partial x^+}{\partial \alpha^j}\right)\right)\,.
\ee
We have to minimize the area
\be\label{eq:Aplane}
\mathcal A=\int dz\, d^{d-2}y\, \sqrt{h}\,.
\ee
We have two equations of motion, one for $x^+$ and one for $x^-$, and the Lagrangian depends only of the derivatives of these fields. The equation of motion for $x^+$ contains only terms proportional to derivatives of $x^-$, and hence can be solved taking 
\be
x^-=0\,,
\ee   
consistently with the boundary condition. This simplifies the equation of motion coming from the variation of $x^-$, since we only need to keep the terms linear in $\partial_i x^-$ in (\ref{eq:Aplane}). The result is
\be
\nabla_y^2 \, x^+ + \frac{1}{f^2} \left(\frac{\partial^2 x^+}{\partial z^2}-\left(\frac{f'}{f}+\frac{d-1}{z}\right) \frac{\partial x^+}{\partial z} \right)=0\,.\label{sss}
\ee
This equation determines the minimal surface. Surprisingly, it is a linear equation for the shape $x^+$. A reason for this is that if $x^+$ is a solution, a scaled $\lambda x^+$ has to be a solution since it arises from boosting. It is the same as the equation for a massless scalar in the bulk metric (\ref{eq:bulk-metric}).

Since we have obtained a minimal surface that lies completely on the $x^-=0$ plane on the bulk, the area on this surface has to be computed with the induced metric 
 \be
ds^2|_{\cal M}=\frac{L^2}{z^2}\left( f^2(z) dz^2 +d\vec{y}^2 \right)\,,
\ee
that is completely independent of the shape of $x^+(z,\vec{y})$. Hence, once we fix a cutoff $z=\epsilon$ and integrate the volume of this $z,\vec{y}$ plane for all $\vec{y}$ and $z>\epsilon$, the area is independent of $\gamma(\vec{y})$. This works for general $f(z)$, i.e., it captures fixed points ($f=1$) and also holographic RG flows. This verifies our arguments in Sec.~\ref{sec:markov}, and leads to the Markov property of the vacuum state in holographic theories. In fact, the area is the same for any surface on the $x^-=0$ plane but only the solution of (\ref{sss}) is extremal.

For pure AdS, we can give an explicit solution for the extremal surface. When $f=1$, (\ref{sss}) reduces to
\be
\left(\nabla_y^2  + \partial_z^2-\frac{d-1}{z} \partial_z\right)x^+=0\,.\label{sss1}
\ee
By Fourier transforming in $\vec{y}$ and choosing the solution regular at infinity, we get the complete solution for the problem
\bea\label{con}
x^+(z,y)&=&\frac{2^{1-d/2}}{\Gamma[d/2]}\int d^{d-2}k\,\, a_{\vec{k}}\,\, e^{i \vec{k}\cdot \vec{y}}\, \,(|\vec{k}| z)^{d/2}\,\,K_{d/2}(|\vec{k}|z)\,, \nonumber\\
a_{\vec{k}}&=&\int \frac{d^{d-2}y}{(2\pi)^{d-2}}\,\, e^{-i \vec{k}\cdot \vec{y}}\, \, \gamma(\vec{y})\,.  
\eea
See also~\cite{Neuenfeld:2018dim}. Eq.~(\ref{sss1}) was also derived in a different context in~\cite{Faulkner:2016mzt}.

\subsection{Regions with boundary on a null cone}\label{subsec:cone}

Next, we consider the entropy of CFTs for regions with boundary on the null cone. One idea would be to obtain the extremal surface and areas by mapping the null plane to the null cone, and then compute the entropy using the metric and a cutoff of fixed $z$ on the cone. We will more simply redo the calculation on the cone directly.  We focus here on smooth curves $\gamma(\Omega)$, and later in Sec.~\ref{subsec:cusps} comment on the effects of cusps.

For pure AdS there is a  conformal transformation from the null plane to the null cone at the boundary that extends as an isometry on the bulk, respecting minimal surfaces and their areas. Hence, the only differences in the computation of the areas in the planar case and the cone can come from the position of the cutoff.  
The isometry of AdS corresponding to (\ref{maa}) is given by extending this conformal transformation to one in a Minkowski space with one more spatial coordinates $z$, and $Z$ respectively. These are just the two bulk coordinates. We have exactly the same formula (\ref{maa}) but where the vectors have now $d+1$ coordinates, and $x^{d+1}=z$, $X^{d+1}=Z$. The AdS metric is invariant under this transformation. The surface $X^0=0$, $X^1=0$, which corresponds to the minimal surface of Rindler space, is mapped to the spherical cup 
\be
|\vec{x}|^2=r^2+z^2=R^2\,,\quad t=-R\,, 
\ee
which is the minimal surface corresponding to the sphere. 

The surface $t+|\vec{x}|=0$, which is the past light-cone in the bulk of the upper tip of the cone, is mapped into the plane $X^-=0$. Then, the minimal surfaces we are interested in will lie on this null cone on the bulk.

To follow the geometric ideas for the Markov property on the original AdS space, we will use the following coordinates
\be
\tilde{r}=|\vec{x}|=\sqrt{r^2+z^2}\,, \hspace{1cm} \tilde{r}^\pm=\tilde{r}\pm t, \hspace{1cm}\tilde{\Omega}\,,
\ee
where $\tilde{\Omega}$ are angular coordinates on the half-sphere $t=\textrm{const}$, $\tilde{r}=\textrm{const}$. For the surface $\tilde{r}^+=0$ each $\tilde{\Omega}$ constant  describes a null line in the bulk having the origin as the future end-point. We will write 
\be
z= \tilde{r} \sin(\theta)\,,\, \theta \in (0,\pi/2)\,,
\ee ç
with $\theta=\pi/2$ corresponding to the point of the sphere further from the AdS boundary, and $\theta=0$ to the boundary. The AdS metric writes
\be
ds^2= L^2\frac{d\tilde{r}^+ d\tilde{r}^- +\tilde{r}^2 d\tilde{\Omega}^2}{\tilde{r}^2 \sin^2 \theta}\,,
\ee
where 
\be
d\tilde{\Omega}^2= d\theta^2+ \cos^2 \theta \;d\Omega_{d-2}^2\,,  
\ee
and $\Omega$ are angular coordinates on a $d-2$ dimensional sphere describing usual polar coordinates in the boundary of AdS.

On the surface $\tilde{r}^+=0$, the induced metric
\be
ds^2=L^2\, \frac{d\tilde{\Omega}^2}{\sin^2 \theta}=L^2\,\frac{d\theta^2+ \cos \theta^2\; d\Omega_{d-2}^2}{\sin^2{\theta}}\,,
\ee
is independent of the remaining coordinate $\tilde{r}^-=2 \tilde{r}=-2 t$. This shows that, if we naively forget about the cutoff, all possible minimal surfaces have the same induced metric and (divergent) area. If we impose a cutoff  on a small $\theta$ independently of $\Omega$ we get again the same result for all minimal surfaces reproducing the previous result for the plane. However, we want to impose a covariant cutoff at fixed $z$ instead. All the dependence on the shape of $\gamma$ will come from this cutoff.

 \subsubsection{Extremal surface and covariant cutoff}

Let us compute the equations for the minimal surface, and check that it lies on $\tilde{r}^+=0$. Writing the $d-1$ coordinates for the sphere described by $\tilde{\Omega}$ as $\alpha^i$ and the sphere metric as $\tilde{g}_{ij}$, we have to extremize the action
\bea
\mathcal A&=&\int d^{d-1}\alpha\,\frac{\det^{1/2}(\tilde{g})}{\sin^{d-1}(\theta)} \,\, \det (\delta^j_l+\tilde{g}^{jk}\partial_k \tilde{r}^+ \partial_l \tilde{r}^-/\tilde{r}^2)^{1/2}\nonumber\\
&=&\int d^{d-2}\Omega\, d\theta\,\frac{(\cos\theta)^{d-2}}{(\sin \theta)^{d-1}}\det (\delta^j_l+\tilde{g}^{jk}\partial_k \tilde{r}^+ \partial_l \tilde{r}^-/\tilde{r}^2)^{1/2}\,,
\eea
with respect to variations of $\tilde{r}^\pm(\tilde{\Omega})$.
The equation of motion for $\t r^-$ is satisfied, along with the boundary conditions, by setting $\tilde{r}^+=0$. The equation of motion of $\tilde{r}^+$ gives
\be\label{eq:EOMr}
\left(\frac{\partial^2}{\partial \theta^2}-\left((d-2) \tan \theta+(d-1) \cot \theta \right)\frac{\partial}{\partial \theta}+\frac{1}{\cos^2\theta}\nabla^2_\Omega \right)(\tilde{r}^-)^{-1}=0\,.
\ee
The same equation holds for $\tilde{r}$ since it is just $\tilde{r}^-/2$. Notice that the equation for $(\tilde{r}^-)^{-1}$ is linear as was the case of $x^+$ for boundaries on the null plane. This is because these two variables are linearly related by the conformal transformation that carries the null plane into the null cone.  

\begin{figure}[h]
\begin{center}
\includegraphics[width=0.4\textwidth]{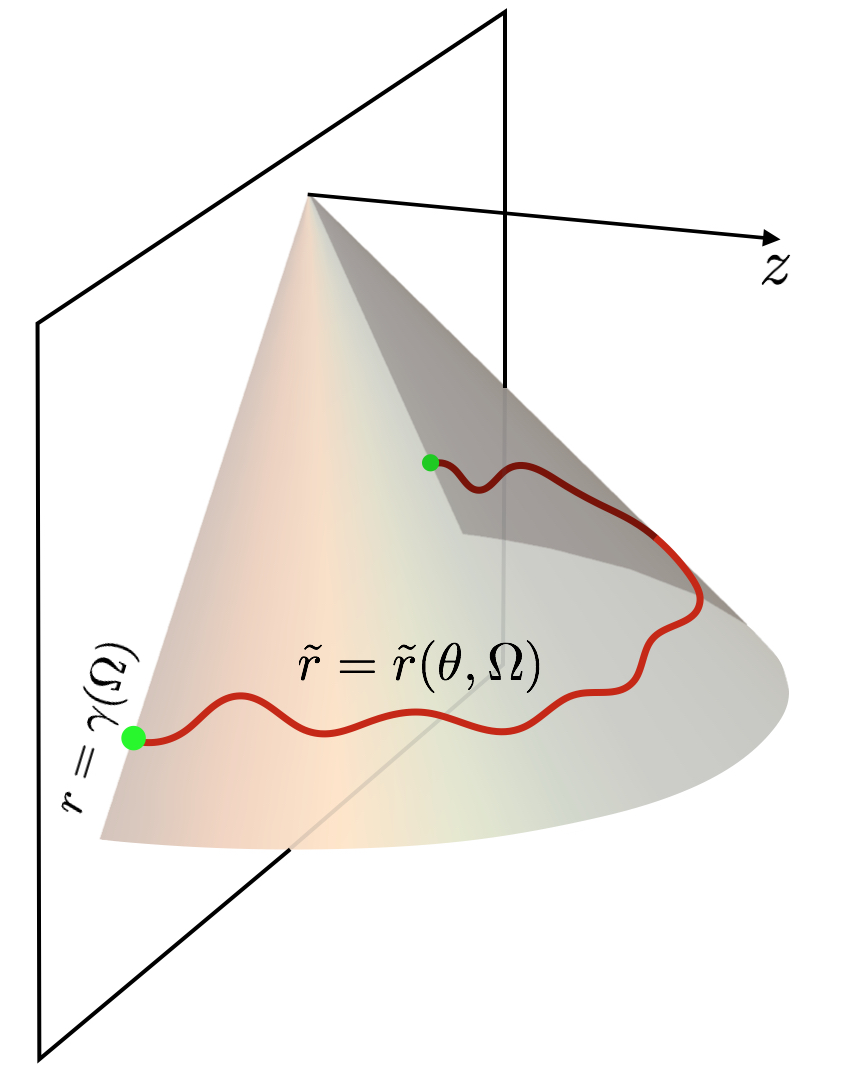} 
\caption{The extremal HRT surface anchored to the locus $r=\gamma(\Omega)$ on a boundary null-cone lies on a bulk null-cone.}
\label{fig:bulknull}
\end{center}
\end{figure}

The boundary curve now is of the form $r=\gamma(\Omega)$, where $r=\sqrt{(x^1)^2+\ldots+(x^{d-1})^2}$. The minimal surface takes the form $\tilde{r}^+=0$, $\tilde{r}(\theta,\Omega)$, with $\tilde{r}(0,\Omega)=r(\Omega)=\gamma(\Omega)$. It lies on the bulk light-cone, as illustrated in Fig.~\ref{fig:bulknull}.

The solution to (\ref{eq:EOMr}) that is regular in the interior $\theta \to \pi/2$ is\footnote{This solution was also obtained in~\cite{Neuenfeld:2018dim}.}
\be\label{eq:solrt}
(\t r(\theta, \Omega))^{-1} = \sum_{n=0}^\infty \sum_I\,\frac{\sqrt{\pi} \Gamma(d-1+n)}{2^{d+n-2}\Gamma(\frac{d}{2})\Gamma(\frac{d-1+2n}{2})}\,a_{nI}\,Y_n^I(\Omega)\,(\cos \theta)^n\,{}_2F_1(\frac{n-1}{2}, \frac{n}{2}, \frac{d-1}{2}+n,\cos^2\theta)\,,
\ee
where $Y_n^I(\Omega)$ are the orthonormal spherical harmonics of degree $n$ on the sphere $S^{d-2}$,
\be\label{eq:laplY}
\nabla_{\Omega}^2 Y_n^I(\Omega)=-(n+d-3)n\,Y_n^I(\Omega)\;,\;n>0\,,
\ee
and $I$ is some multi-index for the eigenfunctions of fixed degree $n$. The prefactor in (\ref{eq:solrt}) is chosen to cancel the value of the hypergeometric function at $\theta=0$, and $a_{nI}$ are the coefficients of the expansion of $\gamma^{-1}$ in spherical harmonics,
\be
\gamma(\Omega)^{-1}= \sum_I\,a_{nI}\, Y_n^I(\Omega)\,.
\ee

We want to impose a standard Lorentz invariant cutoff in 
\be\label{eq:cutoff}
z=\t r(\theta,\Omega) \sin(\theta)=\epsilon\,.
\ee 
Let us denote the solution to this equation by $\theta= \beta(\Omega)$; it will depend on the cutoff $\epsilon$ and on the curve $\gamma(\Omega)$. The minimal area then becomes
\bea
\mc A&=&L^{d-1} \int d^{d-2 }\Omega\,\int_{\beta(\Omega)}^{\pi/2}\,d\theta\,\frac{(\cos \theta)^{d-2}}{(\sin \theta)^{d-1}} \\
&=&L^{d-1}\int d^{d-2 }\Omega\,\frac{1}{d-1}\,(\cos \beta)^{d-1}\,{}_2F_1 (\frac{d-1}{2}, \frac{d}{2},\frac{d+1}{2}, \cos^2 \beta)\,. \nonumber
\eea
This has the form of a local action for the entropy, as in the QFT calculation. Also, as anticipated, all the dependence on $\gamma(\Omega)$ arises through the cutoff $\beta$. Since $\beta \sim \mathcal O(\epsilon)$, we expand in small $\beta$, obtaining
\be\label{eq:As}
\mc A= L^{d-1} \int d^{d-2 }\Omega\, \left \lbrace \frac{1}{d-2} \frac{1}{\beta^{d-2}}- \frac{2d-5}{6(d-4)}  \frac{1}{\beta^{d-4}}+ \left(\frac{3}{8(d-6)}+\frac{d}{18}- \frac{1}{45} \right)\frac{1}{\beta^{d-6}}+ \ldots \right \rbrace + A_0\,.
\ee
Here
\be\label{eq:A0}
A_0 =L^{d-1} \int d^{d-2 }\Omega\,  \frac{\sqrt \pi}{2 \sin \frac{\pi d}{2}} \frac{\Gamma(\frac{d-1}{2})}{\Gamma(\frac{d}{2})}\,= L^{d-1}\,\frac{ \pi^{d/2}}{ \sin \frac{\pi d}{2} \Gamma(\frac{d}{2})}\,. 
\ee

In order to evaluate this expression, we need to solve for $\beta$ in powers of $\epsilon$.
Besides the constant term, (\ref{eq:solrt}) contains a series that starts at order $\theta^2$ and one that starts at $\theta^d$. Explicitly,
\bea
(\t r(\theta, \Omega))^{-1}& =& \gamma(\Omega)^{-1}+ \sum_{n\ge 1,\,I}\,a_{nI} Y_n^I(\Omega)\frac{n(n+d-3)}{2(d-2)} \theta^2 \left\lbrace - 1+ \frac{3 n(n+d-3)-2(d-1)}{12(d-4)} \theta^2 + \ldots \right \rbrace \nonumber \\
&-& \sum_{n\ge 1,\,I}\,a_{nI} Y_n^I(\Omega)\,\theta^d \left \lbrace - \frac{\pi}{2^d \sin \frac{\pi d}{2}} \frac{\Gamma(d+n-1)}{\Gamma(n-1) \Gamma(\frac{d}{2})\Gamma(\frac{d+2}{2})} + \mathcal O(\theta^2)\right \rbrace\,.
\eea
The series in $\theta^2$ can be rewritten in terms of derivatives of $\gamma(\Omega)^{-1}$ by use of (\ref{eq:laplY}),
\bea\label{eq:rtlocal}
(\t r(\theta, \Omega))^{-1}&=&\gamma(\Omega)^{-1}+ \frac{1}{2(d-2)} \nabla_\Omega^2 (\gamma^{-1}) \theta^2 \nonumber\\
&+& \frac{1}{24(d-2)(d-4)} \left(2(d-1) \nabla_\Omega^2(\gamma^{-1}) + 3 \nabla_\Omega^2\nabla_\Omega^2(\gamma^{-1}) \right) \theta^4 + \ldots
\eea
This can also be verified by solving (\ref{eq:EOMr}) in powers of $\theta^2$. In contrast, the series that starts at order $\theta^d$ does not appear to have a local expansion in derivatives of $\gamma^{-1}$. This series is fixed by requiring regularity at the interior $\theta \to \pi/2$, which is the condition that fixed (\ref{eq:solrt}). Such terms end up modifying the EE at order $\epsilon^2$, and hence vanish in the limit in which the UV regulator is taken to zero. We will neglect them in what follows.

Plugging (\ref{eq:rtlocal})  into (\ref{eq:cutoff}) leads to the power-series solution
\be\label{eq:beta}
\beta(\Omega) =\epsilon \gamma(\Omega)^{-1} +\frac{1}{6} \epsilon^3\,\gamma(\Omega)^{-3} \left(1+\frac{3}{d-2} \gamma\,\nabla_\Omega^2\,(\gamma^{-1})\right)+\ldots
\ee
We now use (\ref{eq:As}) and (\ref{eq:beta}) to study the extremal surface area in a derivative expansion. For general $d$, we have
\bea\label{eq:Ad}
\mc A&=&L^{d-1} \int d^{d-2}\Omega\,\Bigg \lbrace \frac{1}{d-2} \frac{\gamma^{d-2}}{\epsilon^{d-2}}- \frac{d-3}{2(d-2)(d-4)}\frac{\gamma^{d-4}}{\epsilon^{d-4}} \left((d-2) + \frac{d-4}{d-3}\gamma \nabla_\Omega^2(\gamma^{-1}) \right)  \nonumber\\
&+&\frac{(d-3)(d-5)}{8(d-2)(d-4)(d-6)}\frac{\gamma^{d-6}}{\epsilon^{d-6}}\Big [ (d-2)(d-4)+\frac{(d-4)(d-6)}{(d-2)(d-3)} (\gamma \nabla_\Omega^2(\gamma^{-1}))^2 \nonumber \\
&-& \frac{d-6}{(d-3)(d-5)}  \left( \gamma \nabla_\Omega^4(\gamma^{-1})-2(d-3)(d-5) \gamma \nabla_\Omega^2(\gamma^{-1})\right)  \Big]+ \ldots \Bigg \rbrace\,.
\eea

\subsubsection{Odd $d$}

For odd $d$, we recognize in (\ref{eq:Ad}) the derivative expansion in terms of  the conformal laplacians presented in (\ref{eq:Sodd}) and
(\ref{eq:Wk}). Furthermore, (\ref{eq:A0}) gives the universal constant term for the EE in holographic theories dual to Einstein gravity. It has the right $(-1)^{\frac{d-1}{2}}$ sign structure. Comparing with (\ref{eq:Sodd}) allows to identify
\be
F=(-1)^{\frac{d-1}{2}} \frac{L^{d-1}}{4 G_N}\,\frac{ \pi^{d/2}}{ \Gamma(\frac{d}{2})}\,. 
\ee
This is the same for any curve $\gamma(\Omega)$ on the cone, and agrees (as it should) with the holographic result for the sphere~\cite{Ryu:2006ef}.\footnote{By a slight abuse of notation, we keep the sign $(-1)^{\frac{d-1}{2}}$ as part of $F$, in agreement with our convention in (\ref{eq:Sodd}). However, the standard notation for $F$ does not include the sign, as in (\ref{eq:EEsphere}).}

In particular, for $d=3$ (\ref{eq:Ad}) becomes
\be
\mc A=L^2 \int d\Omega\,\left ( \frac{\gamma}{\epsilon} -1 + \mathcal O(\epsilon^3) \right)\,.\label{tresd}
\ee
Note from (\ref{eq:Ad}) that the term of order $\epsilon$ is a total derivative $\nabla_\Omega^2(\gamma^{-1})$ in $d=3$. 
For $d=5$, after integration by parts
\be
\mc A=L^4 \int d^3 \Omega\, \left \lbrace \frac{1}{3} \frac{\gamma^3}{\epsilon^3}-\frac{1}{3} \frac{\gamma}{\epsilon}\left(3+ \left(\frac{\nabla_\Omega \gamma}{\gamma} \right)^2 \right)  + \frac{2}{3}+ \mathcal O(\epsilon) \right\rbrace\,.\label{cincod}
\ee
As in (\ref{eq:conformalkin}), the last two terms give the kinetic term for a conformally coupled scalar field, and the first term is a classically conformally invariant potential.

\subsubsection{Even $d$}

For even $d$, the expression (\ref{eq:Ad}) explains the origin of the universal logarithmic terms,
\be
\frac{1}{d-2n} \frac{\gamma^{d-2n}}{\epsilon^{d-2n}}\,\to\, \log \frac{\gamma}{\epsilon}
\ee
for $d=2n$. It also gives rise to the correct WZ terms, although it is not obvious how to rewrite the previous expressions with hypergeometric functions as (\ref{eq:SWZd}). Let us check this for $d=4, 6$.

For $d=4$,
\be
\mc A =L^3 \int d^2 \Omega \left \lbrace  \frac{1}{2} \frac{\gamma^2}{\epsilon^2}- \frac{1}{2} \log \frac{\gamma}{\epsilon} - \frac{1}{4}\left(\frac{\nabla_\Omega \gamma}{\gamma} \right)^2+ \mathcal O(\epsilon^0) \right \rbrace\,.\label{cuatrod}
\ee
The second and third term combine to give the two-dimensional WZ action (\ref{eq:SWZ4}).

For $d=6$,
\bea\label{eq:holo6}
\mc A &=&L^5 \int d^4 \Omega  \Bigg \lbrace  \frac{1}{4} \frac{\gamma^4}{\epsilon^4}- \frac{1}{2} \frac{\gamma^2}{\epsilon^2}\left(\frac{3}{2}+\frac{1}{4}\gamma \nabla_\Omega^2 (\gamma^{-1}) \right) \nonumber\\
&+& \frac{1}{8} \left(3 \log \frac{\gamma}{\epsilon}+ \frac{1}{16}(\gamma \nabla_\Omega^2 (\gamma^{-1}) )^2 - \frac{1}{8} \gamma \nabla_\Omega^4(\gamma^{-1})+\frac{3}{4}\gamma \nabla_\Omega^2 (\gamma^{-1}) \right)+ \mathcal O(\epsilon^0) \Bigg \rbrace\,.
\eea
It is not hard to verify that this result is a linear combination of the WZ action (\ref{eq:SWZ4new}) and the two invariant terms that obtain from $\hat R^2$ and $\hat R_{ab}^2$ in (\ref{eq:Snodd}). This is a nontrivial check, given that the four terms in the last line of
(\ref{eq:holo6}) are reproduced in terms of the QFT formula that has three independent contributions at this order.

\subsection{Comments on cusps}\label{subsec:cusps}

The holographic formula for the entropy contains terms depending on derivatives of $\gamma$. Here we want to comment on the interpretation of these terms when $\gamma$ is not smooth. We will only treat the case of a cusp, that is, the case of a jump in derivatives, and for simplicity will keep the discussion centered in low dimensions $d=3,4$.  

For a smooth surface,  $\nabla_\Omega^2 (\t r^{-1})$ is finite as $\theta \to 0$; then we found in (\ref{eq:rtlocal}) that $\partial_\theta (\t r(0,\Omega)^{-1}=0$ and our previous results apply. However, this need not be true near a cusp. Before getting to the cusps, let us assume that there is some power-law singularity as we approach the boundary,
\be
\nabla_\Omega^2 (\t r^{-1}) = C_0 \,\theta^{-\nu}\;,\;\theta \to 0\,.
\ee
Solving the equation of motion for small $\theta$ then gives
\be
\t r^{-1} \approx \frac{C_0}{(2-\nu)(d+\nu-2)}\,\theta^{2-\nu} \,.
\ee
Therefore, negative powers of $\theta$ from $\nabla_\Omega^2 \t r^{-1}$ will indeed modify the expansion (\ref{eq:rtlocal}). We will now see that $\nu=1$ at codimension one cusps.

For simplicity, let us focus on $d=3$, and consider a cusp at $\phi=\phi_0$ with local angle $\alpha$. Then, close to the cusp, $\gamma''(\phi) \sim \delta(\phi-\phi_0) \tan \alpha$. At finite $\theta$, this delta function is smoothed;  we should recover an approximant of the delta function as $\theta \to 0$. By dimensional analysis,
\be\label{eq:rcusp}
\partial_\phi^2(\t r(\theta,\phi)^{-1}) \approx \frac{\tan\alpha}{\pi} \,\frac{\theta}{\theta^2+(\phi-\phi_0)^2}\,,
\ee
valid for small $\theta$ and near the cusp. Indeed, it is not hard to check that
\be
\lim_{\theta \to 0} \,\frac{1}{\pi}\,\frac{\theta}{\theta^2+(\phi-\phi_0)^2}= \delta(\phi-\phi_0)\,.
\ee

Plugging (\ref{eq:rcusp}) into the minimal area equation and expanding for small $\theta$, we find
\be\label{eq:drcusp}
\partial_\theta(\t r(0,\phi)^{-1})= \left \lbrace \begin{matrix}\frac{1}{2\pi}\tan \alpha\,, & \phi = \phi_0 \\ 0 \,,& \phi \neq \phi_0 \end{matrix} \right.
\ee
This can also be checked by computing the Fourier coefficients and performing the full sum (\ref{eq:solrt}). For instance, the calculation can be done explicitly for a cusp of the form $\sin |\phi|$.

The same will happen for $d\ge 4$ as long as the cusp has codimension one, with $\phi$ above playing the role of the local normal coordinate. Indeed, for a cusp at $\phi_0$ that locally looks like $\gamma^{-1}\sim |\phi-\phi_0|$, we have $\nabla^2_\Omega \gamma^{-1} \sim \delta(\phi-\phi_0)$; this is just the familiar fact that $|\phi-\phi_0|$ is the one-dimensional Green's function. This also says that contributions from cusps of higher codimension will be smaller. Indeed, to get a delta function from $\nabla^2_\Omega \gamma^{-1}$ at codimension $n$, we need $\gamma^{-1} \sim 1/|\vec x-\vec x_0|^{n-2}$. However, we are considering curves without such divergences, and so all the cusp contributions will have $\nu<1$, with $\nu=1$ for codimension one cusps only.

We conclude that the area integral is not affected by null cusps, since (\ref{eq:drcusp}) modifies the expansion of $\beta(\Omega)$ on a measure zero set of points (the cusps).
Therefore the formula (\ref{eq:Ad}) for the entropy has to be integrated on each side of the cusp where the regular expansion in $\theta$ works, without any further cusp contribution. In consequence, the Markov property continues to hold when there are cusps. 

However, we cannot eliminate boundary terms in the integration by parts when there is a cusp.    
 For example, the finite term with a Laplacian in $d=4$ can be treated in the following way when there are cusps. We integrate in the smooth patches $P_i$ to get
\be
 \int_{P_i} d\Omega\, r \nabla^2_\Omega r^{-1}=\int_{P_i} d\Omega\, \frac{\nabla_\Omega r \cdot \nabla_\Omega r}{r^2}-\int_{\partial P_i} dl\, \eta\cdot\frac{ \nabla_\Omega r}{r}\,,\label{boundarys}
\ee
where the scalar products are with the sphere metric, and $\eta$ in the last term is the outward pointing unit normal to the boundary $\partial P_i$ on  the sphere. The first term has a discontinuous but bounded integrand on the boundary (the position of the cusp). 

It is interesting to see that written in this way, the contributions of the local integrand cancel locally in the SSA relation, but the second term will cancel in the SSA relation because it has opposite contributions to the intersection and the union. This is because these have locally the same $(\nabla_\Omega r)/r$ at the points of the boundary of the patch, but opposite $\eta$.

\subsection{Higher derivative gravity theories}

In the remaining of this section, we will extend the previous results to include stringy and quantum effects. 

Higher derivative gravity theories in the bulk around an AdS solution represent different CFTs incorporating $1/\lambda$ corrections, with $\lambda$ the t'Hooft coupling. A general form of the EE functional corresponding to higher derivative Lagrangians was discussed in \cite{Dong:2013qoa,Camps:2013zua}. The result is a geometric functional computed on the generalized Ryu-Takayanagi surface $\Sigma$, including curvature and extrinsic curvature corrections. 
 Here we want to briefly discuss how the main results of the preceding sections are expected to remain unchanged for these models. 

For a gravity action that is a function of the curvature tensor, the generalized entropy functional has two types of terms. The first is  Wald's entropy formula
\be
-2 \pi \int d^{d-1}y\, \sqrt{g}\, \frac{\partial L}{\partial R_{\mu\rho\nu\sigma}} \varepsilon_{\mu\rho}\varepsilon_{\nu\sigma}\,, \label{wald}
\ee 
where 
\be
\varepsilon_{\mu\nu}= n_\mu^{(a)} n^{(b)}_\nu \varepsilon_{ab}\,,
\ee
the vectors $n^{(a)}$, $a=1,2$, are two normalized vectors normal to the codimension two surface, and $\varepsilon_{ab}$ is the usual two-dimensional Levi-Civita tensor. In what follows we find it convenient to choose $n^{(a)}$ as two null vectors orthogonal to the surface, normalized by $n^{(1)}\cdot n^{(2)}=1$.  
The second type of terms involves the extrinsic curvatures of the surface and is proportional to
\bea
&&\int d^{d-1}y\, \sqrt{g}\,\frac{\partial^2 L}{\partial R_{\mu_1\rho_1\nu_1 \sigma_1}\partial R_{\mu_2\rho_2\nu_2 \sigma_2}} \,K_{\lambda_1\rho_1\sigma_1} \,K_{\lambda_2\rho_2\sigma_2}\label{otrooo}\\
 &&\nonumber\hspace{2cm} \times \left((\eta_{\mu_1\mu_2} \eta_{\nu_1\nu_2}-\varepsilon_{\mu_1\mu_2}\varepsilon_{\nu_1\nu_2})\eta^{\lambda_1\lambda_2}+(\eta_{\mu_1\mu_2}\varepsilon_{\nu_1\nu_2}+\varepsilon_{\mu_1\mu_2}\eta_{\nu_1\nu_2})\varepsilon^{\lambda_1\lambda_2}\right)\,.
\eea
Here $\eta$ is the projector onto the vector space normal to the surface
\be
\eta_{\mu\nu}=n^{(1)}_\mu n^{(2)}_\nu+n^{(2)}_\mu n^{(1)}_\nu\,.
\ee
The extrinsic curvature is given by
\be
K_{\lambda\mu\nu}= n^{(2)}_\lambda P^\alpha_\mu P^\beta_\nu \nabla_\alpha n^{(1)}_\beta + n^{(1)}_\lambda P^\alpha_\mu P^\beta_\nu \nabla_\alpha n^{(2)}_\beta\,,\label{ex}
\ee
where $P$ is the projector to the tangent space of the surface
\be
P^\alpha_\mu= g^\alpha_\nu-\eta^\alpha_\mu\,. 
\ee

The bulk metric is pure AdS corresponding to vacuum CFT.
In AdS the curvature tensor is proportional to combinations of product of the metric tensor. In consequence, Wald's term (\ref{wald}) is proportional to the area functional. 
 
Let us consider a surface $\Sigma$ that lies on the bulk null cone $\tilde{r}^+=0$. In that case we can choose $n^{(1)}$ to be the Killing null vector parallel to the cone. Then we have
\be
 (\nabla_\alpha n^{(1)}_\beta+\nabla_\beta n^{(1)}_\alpha)  =0\,.   
\ee
As the extrinsic curvature tensor (\ref{ex}) is symmetric in $\mu,\nu$ the contribution of the derivative of $n^{(1)}$ vanishes.
In consequence only one term remains in the extrinsic curvature (\ref{ex}) and the integrand in (\ref{otrooo}) vanishes as well. In addition,  
 we have here a situation analogous to the one of surfaces $\gamma$ in a null plane discussed in Sec.~\ref{sec:markov}. The areas of any two surfaces lying on this null cone in AdS are equal since only the projection of the surface orthogonal to $n^{(1)}$ contributes, and there is an isometry that shows that these projections are equal along  the direction of the null ray. Then, on the null cone in the bulk, all surfaces give the same value of the functional. 
 
The equations that fix the position of $\Sigma$ in the general case follow by extremizing the entropy functional \cite{Dong:2017xht}. For surfaces on the null cone, the variations of the entropy functional for variations of position also contained in the null cone, vanish. Hence, analogously to the case of Einstein gravity treated above, one of the equations of motion is solved precisely by placing $\Sigma$ on the null cone, and this is compatible with the boundary conditions. The other equation of motion will fix the shape of the surface on the cone itself. On the cone, the functional is just proportional to the area, but this need not be the case for deformations that take the surface outside the cone. Hence, we expect the differential equation for $\tilde{r}^-$ to get modified by the higher derivative terms in the Lagrangian. However, this equation should still be linear. This is because, as we have explained in section \ref{subsec:plane}, boost invariance will lead to a linear equation for regions on the null plane on the boundary, and a conformal transformation will give a  linear equation for $(\tilde{r}^-)^{-1}$.   

In any case, once the surface is determined, the Markov property follows from the fact that the functional on the cone reduces to a term proportional to the area, and the area on the cone is independent of shape. Then, the result can only be affected by the position of the cutoff. Again, we will have a local expression for the entropy as a function of $\gamma$, with the same types of terms found in Sec.~\ref{sec:qft}. The only change can be in the coefficients of the independent terms, in particular the value of the anomaly. This can be calibrated by computing the entropy of the sphere. See for example \cite{Hung:2011xb}.   

\subsection{$1/N$ corrections}

According to \cite{Faulkner:2013ana},  $1/N$ corrections to the entanglement entropy in the large $N$ limit come from quantum corrections in the bulk. One has to add to the holographic entropy the entanglement entropy of quantum fields living in the bulk across the Ryu-Takayanagi surface. 

For the regions on the light-cone we are considering, the entangling surfaces all lie on the bulk light-cone $\tilde{r}^+=0$ in AdS. Then, we can apply an argument analogous to the one on Sec.~\ref{sec:markov} for the null plane in Minkowski space. The bulk EE has to be a functional of surfaces on the light-cone, and this light-cone is mapped into itself by isometries of AdS which correspond to conformal symmetries of the boundary theory. For example, we can take a surface $\gamma$ on the boundary, and a sphere $\gamma'$ on the light-cone which does not cut $\gamma$. The modular flow corresponding to $\gamma'$ will move $\gamma$ towards $\gamma'$ as much as we want. In the bulk, this corresponds to an isometry that will squeeze as much as we want the entangling surface of $\gamma$ towards the entangling surface of the sphere $\gamma'$ (which is a sphere in the bulk). This symmetry keeps the vacuum invariant and respects a covariant cutoff in the bulk. Hence it will keep the bulk EE invariant. 

We conclude that quantum corrections in the bulk, except for terms coming from the UV cutoff of the boundary theory, will be the same for all regions on the light-cone, and will not spoil the Markov property. We expect the same structure of the entropy as in Sec.~\ref{sec:qft}, with some corrections in the different coefficients for the independent possible terms.          
 
\section{Revisiting the entropic proof of the $a$-theorem}
\label{sec:athm} 

In the previous sections we obtained the explicit form of the CFT entropy on the null cone and worked out the holographic case. In this section we will use this information to check the arguments leading to a proof of the $a$-theorem in $d=4$ in~\cite{Casini:2017vbe}.  These followed the lower dimensional cases ($d=2,3$) treated in~\cite{Casini:2004bw,Casini:2012ei}, where the strong subadditive property of the entropy was used for spheres (intervals or circles in $d=2$ and $d=3$ respectively) on the light-cone to show the monotonicity of the $c$ and $F$ quantities. In particular, the result (\ref{eq:Seven}) for the entropy for arbitrary regions on the null cone will allow us to see explicitly why the Markov property has to be invoked as a key ingredient in $d=4$, as opposed to the $d=2$ and $d=3$ cases. However, from the outset we can say that the Markov property plays an important hidden role even in dimensions lower than $d=4$. This is because if the  strong subadditive inequality can teach us something non-trivial about the RG running, it must be the case that this inequality saturates for a CFT, where no relevant RG running is taking place. This shows the precise reason of the geometric setup of these theorems involving regions on the null cone. This is basically the only case where the Markov property holds for a CFT.\footnote{For regions $A$ and $B$ where $A-B$ and $B-A$ contain non-trivial spacial slices the Markov property cannot hold since there is quantum entanglement between them, as can be seen from the failure of Bell's inequalities for the correlators \cite{Verch:2004vj}.} 

\begin{figure}[t]
\begin{center}
\includegraphics[width=0.45\textwidth]{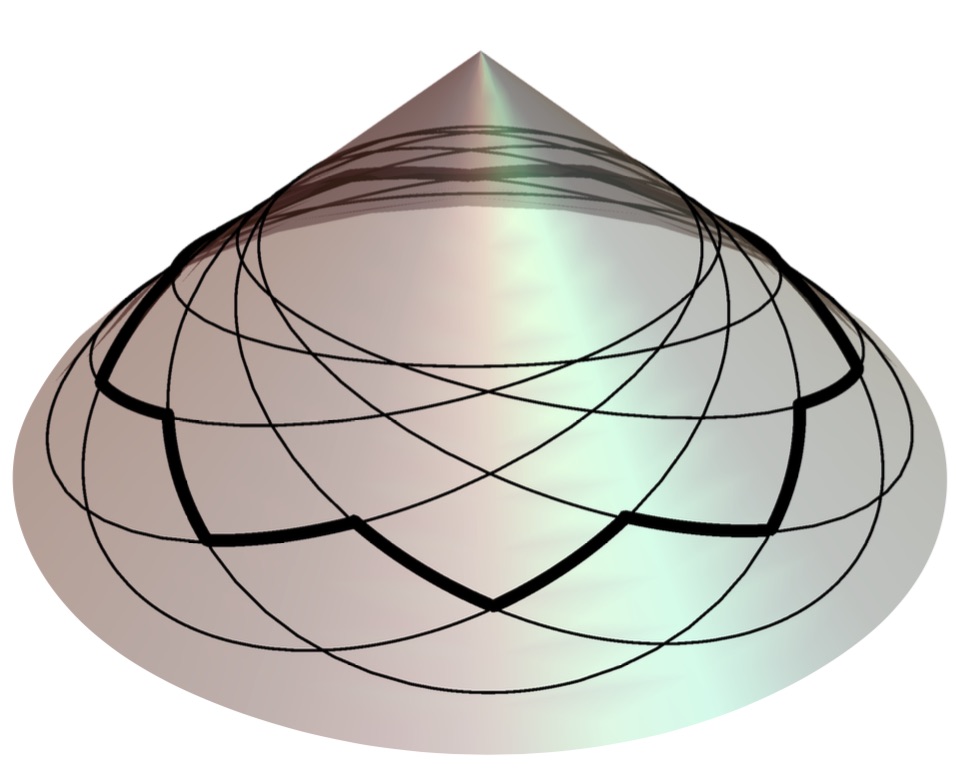} 
\caption{Boosted circles lying on the null cone in $d=3$. The vertical axis of the cone gives the time direction.}
\label{fig:boosted}
\end{center}
\end{figure}

Let us first review the arguments in~\cite{Casini:2012ei}. 
We start with a boosted sphere of radius $\sqrt{r R}$ lying on the  null cone between the time slices at time $|t|=r$ and $|t|=R>r$. We then take a large number $N$ of rotated copies of this sphere, as equally distributed on the unit sphere of directions as possible.\footnote{It is not possible to distribute them in a regular fashion for $d>3$. The details of this distribution on the unit sphere of directions turns out to be irrelevant as far as a uniform distribution is approached for large $N$.} From strong subadditivity we get the inequality in the limit of large $N$
\begin{equation}
S(\sqrt{r R})\ge  \int_r^R dl\ \beta(l) \tilde{S}(l)\,. \label{41}
\end{equation}
In this expression $ \tilde{S}(l)$ are the entropies of ``wiggly'' spheres that come about in the process of intersecting and joining boosted spheres in the SSA inequality -- see Fig.~\ref{fig:boosted}. The wiggly spheres have an approximate radius $l\in (r,R)$, and lie around the surface of equal time $|t|=l$; the deviations from the perfect sphere of radius $l$ at $|t|=l$ form the wiggles, that lie on the null cone, and have a typical width $\sim l/N^{1/(d-2)}$ that tends to zero for large $N$.  
$\beta(l)$ is the density of wiggly spheres as the number of boosted spheres $N\rightarrow \infty$, divided by $N$.\footnote{Strictly speaking the integral in (\ref{41}) is a sum over $N$ wiggly sphere entropies divided by $N$. The notation with an integral and a density of wiggly spheres of the same radius is a convenience here, that will make sense for later expressions when we take the limit $N\rightarrow \infty$, and more information about the entropies of the wiggly spheres is introduced.} It is given by    
\begin{equation}
\beta(l)=\frac{\text{Vol}(S_{d-3})}{\text{Vol}(S_{d-2})}\,  \frac{2^{d-3} (r R)^{\frac{d-2}{2}} \left( (l-r)(R-l) \right)^{\frac{d-4}{2}}    }{ l^{d-2}  (R-r)^{d-3}}\,,
\end{equation}
 normalized to have unit integral, 
\be
\int_r^R dl\, \beta(l)=1\,.\label{normali}
\ee

In a sense these wiggly regions tend to spheres of radius $l$ for large $N$, but we have to work out how exactly the entropies behave in this limit. Note that even if the amplitude of the wiggles decreases with $N$ this is not the case for their slope, which remains a fixed function of $l$ in the limit $N\rightarrow \infty$.

At this point three different questions arise which have to be understood in order to extract useful information for the monotonicity theorems from (\ref{41}).
 The first question is if this inequality contains cutoff independent information, that is, if the divergent terms cancel between the two sides of the inequality. Since divergences are local on the boundary of the regions this can be rephrased as if the new features on the wiggly spheres, coming from the locus of intersections of two or more spheres for example, gives place to new unbalanced divergent terms or not. The second question is whether, in case the inequality contains information about finite quantities, this can be extracted in a useful way. In other words, whether the wiggly sphere entropies can be related to sphere entropies. The third and last question is if the inequality will teach us something about the central charges at the fixed points of the RG. We will discuss these three questions in turn.  
 
\subsection{The inequality is UV finite}

Unbalanced divergences in the inequality in principle could appear due to the cusps formed at the intersection and union of smooth spheres. We want to present a slightly different geometrical setup which bypasses this issue about divergent terms in any dimension. 

\begin{figure}[t]
\begin{center}  
\includegraphics[width=0.8\textwidth]{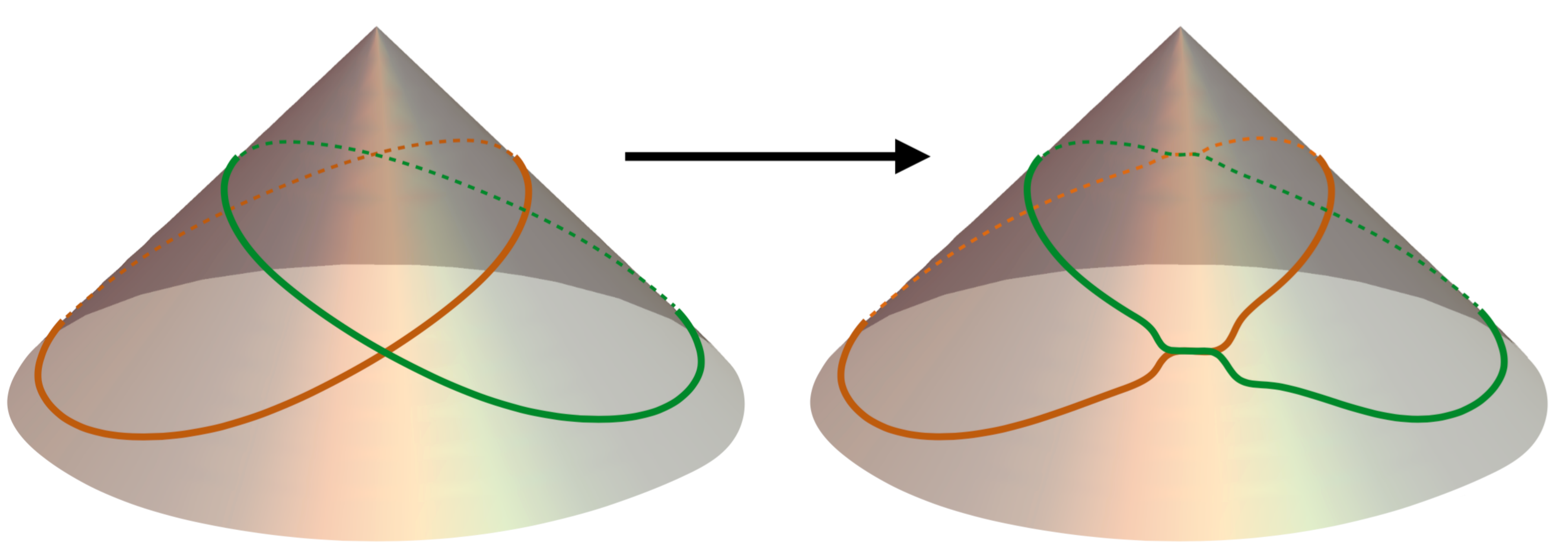}
\captionsetup{width=0.9\textwidth}
\caption{Deformations of spheres to smooth out intersections and unions on the light-cone.}
\label{fig:smooth}
\end{center}  
\end{figure}

The idea is to slightly deform the spheres of radius $\sqrt{r R}$ on the left hand side of the inequality along the null cone and around the points of intersection with other rotated spheres such that all intersections and unions are now smooth (we can choose infinitely many smooth derivatives). See Fig.~\ref{fig:smooth}. In this case there are no cusps and it is clear that the divergent terms cancel in any regularization. The price we pay is that now we do not have perfect spheres on the left hand side of the inequality, and they are replaced by wiggly spheres of approximate radius $\sqrt{r R}$. The inequality now reads
\begin{equation}
\frac{1}{N}\sum_i \tilde{S}_i(\sqrt{r R})\ge  \int_r^R dl\ \beta(l) \tilde{S}(l)\,, \label{441}
\end{equation}
where $\tilde{S}(l)$ is the entropy of a wiggly sphere of approximate radius $l$ and again the integral on the right hand side is a shortcut for a sum over $N$ terms. In the present case this is not a big price to pay since we already have to deal with the wiggly spheres on the right hand side. The size of the new wiggles used to smooth out the cusps can be made arbitrarily small.    

While this approach sidesteps the issue of divergences arising at the cusps, in \cite{Casini:2017vbe} we argued that the divergences cancel out from (\ref{41}), even in presence of cusps. We argued in two steps, assuming a covariant cutoff.\footnote{A general definition of a covariant cutoff for an arbitrary QFT can be provided using mutual information along the same lines as has been done for $d=3$ in \cite{Casini:2015woa}. This is reviewed in Appendix \ref{Appendix:mutual}.} For completeness, in the rest of this section we will review and discuss these arguments.

\bigskip

1) First, since (\ref{41}) was obtained by a series of SSA inequalities, the Markov property requires that the divergences cancel for a CFT. Let us see how this comes about.  
The new divergences on the new local features of the intersections and unions are given by integrals of local geometric terms on the defects of the surface. An essential point is that these defects live on a null cone.  The leading divergence is proportional to the defect dimensions, and we also have new terms for all subleading integer powers corresponding to integration of the defect curvatures along the defect. For a CFT the dimensions of these terms are compensated by negative integer powers of the cutoff $\epsilon$ (or a logarithm if the power is zero). 

Let us focus on $d=4$.  We have linear terms growing as $L/\epsilon$ from the intersection of two spheres in a curve of size $L$, and from the same defect, a term proportional to $\log (L/\epsilon)$ due to the integral of the curvature of the intersection curve along the defect. From the vertex of the intersection of three spheres we should also get a logarithmic term. 

Now, the argument is that the coefficients of these contributions are either zero or have opposite sign for the contributions of the defect to the union and the intersection that gave place to it in the SSA inequality. Let us first consider the leading divergences, where no curvature terms are present. Hence the contribution is the same as for the same type of defect on a null plane rather than a cone. The defect will not contribute because there is no geometric quantity depending on the defect ``angles'' on which the entropy can depend making the defect contribution different from the plane without defect. These is just a manifestation of the argument in Sec.~\ref{sec:markov} about functionals on a null plane being independent of $\gamma$. In other terms, boosting these geometries while keeping the null plane and the location of the defect invariant, one can squash the planes and make them as similar to a single plane without defect as we want. To be more explicit, take for example the case of the vertex in $d=4$. The vertex defines three spatial lines with unit tangents $t_1$, $t_2$ and $t_3$. However, these tangents live in a three-dimensional null plane. Therefore they all can be written as linear combinations of a spatial vector living in a two dimensional plane orthogonal to the null vector $k$ and $k$ itself, $t_i=v_i+\alpha_i k$, with $v_i^2=1, v_i\cdot k=0$. In any invariant formed by the three vectors all contributions from the component along $k$ will vanish and then the invariant will be the same as the one formed by three lines in a single two dimensional plane, which of course does not define a real vertex. 
 
Hence we conclude that these terms have zero coefficient and do not appear in the entropy. 
The holographic examples in Sec.~\ref{sec:holo} also illustrate this. For $d=3$ and $d=4$ we showed there is no $\log(\epsilon)$ (resp. no $1/ \epsilon$) contribution from the cusps.    

In $d=4$ we also have the possibility of a curvature term on the intersection of two spheres. This can sense the form of the null cone and in this way bypass the arguments in Sec.~\ref{sec:markov}. 
In writing the contribution of the curvature term we are allowed to use the gradient operator $\nabla_\mu$ on the vector $k$ for example, to produce local invariants. However, these gradients are defined on the defect only, and then the indices of the derivatives have to be contracted with one of the defect directions. This defect is locally formed by the intersection  of two spatial planes inside the same null hyperplane with null vector $k$. Each spatial plane has another null vector $q_i$ that defines it, such that $q_i^2=0$, $q_i\cdot k=1$. There is an ambiguity in this representation of the planes in the scale of $k$, as we can freely rescale $k\rightarrow \lambda k$, $q_i\rightarrow (1/\lambda) q_i$. Then, in order to produce the integrand of the contribution we have to write an invariant using the same number of vectors $q_i$ than of $k$. The only non trivial invariant with the right dimensions is
\be
\int dx^\mu \, (\nabla_\mu k^\alpha) \,k^\beta \,q_1^\gamma \,q_2^\delta\, \varepsilon_{\alpha\beta\gamma\delta}\,. 
\ee
This requires a choice of ordering of the two vectors $q_1$, $q_2$, which can be assigned for example choosing first the one to the right of the direction of integration along the intersection. This orientation changes sign when we compute the contributions of this defect to the intersection and the union of the two spheres, and hence the full $\log \epsilon$ contribution of these defects to the SSA inequality vanish. 

In our general analysis in Sec.~\ref{sec:qft}, and the holographic case in Sec.~\ref{sec:holo}, we have in fact learned a bit more. We have shown that the total coefficient of the $\log \epsilon$ term is a topological invariant and it is always the same for any shape on the null cone. This is given by an integral of the intrinsic curvature  of the surface, giving the Euler number (the only non vanishing term in Solodukhin's formula~\cite{Solodukhin:2008dh} in this case). Hence, the $\log(\epsilon)$ contribution clearly cancels from SSA. To see how this fits with the previous argument, suppose we have a normalized contribution $\log(\epsilon)$ for any shape and we are doing the SSA of two spheres of radius $\sqrt{r R}$. The logarithmic coefficient for the intersection and union should be of the form 
\bea
1=\frac{\textrm{area}_\cap}{4\pi r R} + \textrm{cusp}_\cap \,,\\
1=\frac{\textrm{area}_\cup}{4\pi r R} + \textrm{cusp}_\cup \,,
\eea
where the first term on the right hand side comes from integration of the constant intrinsic curvature of the spheres and is proportional to the total solid angle. Summing these two equations and using   $\frac{\textrm{area}_\cap+\textrm{area}_\cup}{4\pi r R} =2$ we get $\textrm{cusp}_\cap=-\textrm{cusp}_\cup$, which coincides with the previous argument. 

\bigskip

2) The previous argument  shows that the inequality is free from divergences for a CFT. If we add a relevant deformation other divergent terms can appear with different powers of $\epsilon$, and where some cutoff powers are replaced by powers of the coupling constant. However, the important point is that these terms are again local on the boundary and have to have the same geometric structure as for a CFT, being integrals of local geometric tensors on the boundary. That is, the only change is in replacements of the cutoff by coupling constants. Then, the previous argument still gives an inequality free of divergences.


\subsection{Converting wiggly spheres into spheres}

We would like to convert wiggly spheres into spheres in (\ref{41}) or (\ref{441}). It turns out that this is correct for $d=2$ (since there are no wiggly intervals) and for $d=3$, where terms produced by the wiggles go to zero for large $N$. This is not the case for $d=4$, and the naive replacement of wiggly spheres by spheres just violates the Markov property. Let us see this in more detail. 

For a CFT in $d=4$ the entropy for a sphere has the form
\be
S(l)=c\, \frac{l^2}{\epsilon^2}-4 a \log(l/\epsilon)\,.
\ee
If we attempt to plug this formula into the Markov equation, assuming wiggly spheres can be replaced by spheres, 
\begin{equation}
S(\sqrt{r R})=  \int_r^R dl\ \beta(l) S(l)\,, \label{4441}
\end{equation}
we find this is not correct. The area term does indeed cancel since
\be
(\sqrt{r R})^{d-2}= \int_r^R dl\ \beta(l) l^{d-2}\,,\label{acan}
\ee
and the constant $\log(\epsilon)$ term cancels as well due to (\ref{normali}). However, this is not the case for the $-a \log(l)$ term. 

The issue here is that there is a nontrivial contribution to the wiggly sphere entropy from the finite term in (\ref{cuatrod}) that comes together with the logarithmic term; this contribution, however, cancels for spheres at constant $t$ on the right hand side of (\ref{4441}). This invalidates the replacement of wiggly spheres by spheres. We will now see that taking this difference into account correctly restores the Markov equality.

With $l=\sqrt{x^2+y^2+z^2}$, and $\theta$ the usual polar angle, the equation for the boosted sphere of radius $\sqrt{r R}$ is 
\be
|t|=l=\frac{2 r R}{r+R-(R-r)\cos(\theta)}\,.
\ee
We have 
\be
\frac{1}{2}\frac{(\nabla_\Omega \gamma)^2}{\gamma^2}= \frac{1}{2}\left(\frac{1}{l}\partial_\theta l\right)^2=\frac{(R-l)(l-r)}{2 r R}\,. 
\ee
We get a constant integrand (except for higher order terms in $1/N$) on the surface of the wiggly sphere of approximate radius $l$.\footnote{The boundary terms in (\ref{boundarys})  cancel automatically in the sum over wiggly spheres.} 
Taking into account this term, the Markov equation for the finite terms
\be
\log(\sqrt{rR})=\int_r^R dl\, \beta(l)\,\left(\log(l)+\frac{(R-l)(l-r)}{2 r R}\right)\,,
\ee
is now satisfied, once we replace $\beta=\frac{r R}{l^2 (R-r)}$ corresponding to $d=4$. Note that the cancellation happens in each SSA equality but in terms of the wiggly spheres it happens ``non locally'', and takes all the range $l\in (r,R)$.

Therefore, a finite term coming from the wiggles obstructs replacing the wiggly spheres by spheres. The idea of~\cite{Casini:2017vbe} was to take advantage of the Markov property of a CFT to subtract from the inequality for the entropies $S$ of the deformed theory the equation corresponding to the entropies $S_0$ of the UV CFT. This can be done at no cost since the SSA of $S_0$ vanishes exactly. We have shown that, in addition, the divergent terms coming from massive deformations are also Markovian and cancel in the SSA inequality; we can subtract them as well, without spoiling the inequality.     
Then, in any dimensions,  we safely replace 
\be
S(l)\rightarrow \Delta S(l)=S(l)-S_0(l)-\textrm{massive divergent terms}\,,
\ee
 in (\ref{441}). Now the finite terms of the wiggles coming from the UV fixed point disappear in the subtraction, and we are free to replace subtracted wiggly spheres by subtracted spheres, taking the limit $N\rightarrow \infty$, and getting the inequality
  \begin{equation}
\Delta S(\sqrt{r R}) \ge   \int_r^R dl\ \beta(l) \Delta S(l)\,.  \label{111}
\end{equation}
 
We still have to check that there are no finite terms induced by a mass parameter that give a contribution for the wiggles that survive in the limit of small wiggles for the deformed theory. In fact, the difference in the EE from a wiggly and non wiggly sphere is controlled by the UV. These terms should be proportional to some mass scale of the square coupling constant $g^2$ of the theory deformation at the UV, which must be compensated by powers of $r$ and positive powers of the distance scale set by the wiggles size. In consequence, they do not contribute in the large $N$ limit. In more detail, a local term should be of the same form as the ones encountered for CFTs but where a power of the cutoff has been replaced by one of a mass parameter. These contributions are divergent except for some non generic perturbation dimensions. In any case a local term is always Markovian and can be subtracted as well. If the term induced by the deformation is non local,\footnote{See for example eq. (\ref{oyo1}) in the next subsection.} then the change from the wiggly sphere to the sphere is suppressed by powers of the wiggly size, and does not contribute in the limit.   
We have computed these wiggly massive corrections holographically in Appendix \ref{app:RG}. The result agrees with these expectations.

Note that for $d=3$ the formula (\ref{tresd}) gives no contribution for the wiggles, and we can safely replace wiggly circles by circles without subtracting the CFT entropies. But this is not the case in higher dimensions. 
  
\subsection{Irreversibility theorems}

We then have (\ref{111}) for spheres in any dimension, where the UV CFT entropy along with other possible divergent contributions have been subtracted. These inequalities are equivalent to the differential ones obtained taking the limit $r\rightarrow R$:
\be
r\, \Delta S''(r) -(d-3) \Delta S'(r)\le 0\,. \label{tera}
\ee 
Writing the entropy as a function of the area $a$ rather than the radius, we get the compact expression
\be
\Delta S''(a)\le 0 \label{sisi}
\ee 
valid in any dimension. Thus the constraint for $\Delta S$ is that it must be concave as a function of the area.

For completeness, let us briefly review here the results of~\cite{Casini:2017vbe}. With our definition of $\Delta S$, that has the entropy with the UV CFT terms and other possible divergent terms subtracted, in the UV limit of small $r$ all local geometric terms vanish and we get the leading ``nonlocal'' term (see e.g.~\cite{Metlitski:2011pr, Liu:2012eea, Liu:2013una} for the structure of the entropy of spheres at fixed points)
\be
\Delta S_{UV}(r) \sim  c_0 \,g^2 r^{2(d-\Delta)}+\ldots =c_0\, g^2 a^{\frac{2(d-\Delta)}{d-2}}+\ldots\,,\label{oyo1}
\ee
where the ellipsis are higher powers in $r$. In the IR fixed point all contributions (except the universal term) are local (proportional to integral of curvatures on the surface) and we have
\bea 
\Delta S_{IR}(r)&=&\Delta \mu_{d-2}\,r^{d-2}+\Delta \mu_{d-4}\, r^{d-4}+\ldots  + \left\lbrace \begin{array}{l} (-)^{\frac{d-2}{2}} 4\,\Delta A\, \log(m R)\,\, d \,\, \textrm{even}\\ (-)^{\frac{d-1}{2}} \Delta F \hspace{1.9cm}d\,\,\textrm{odd }  \end{array}\right.\,\\
\nonumber
&=&\Delta \mu_{d-2}\, a +\Delta \mu_{d-4}\, a^{\frac{d-4}{d-2}}+\ldots  
  + \left\lbrace \begin{array}{l} \frac{(-)^{\frac{d-2}{2}} 4}{(d-2)}\Delta A\, \log(m^{d-2}a)\,\, d \,\, \textrm{even}\\ (-)^{\frac{d-1}{2}} \Delta F \hspace{1.9cm}d\,\,\textrm{odd }  \end{array}\right.\,,
\label{even1}
\eea
with $m$ a characteristic energy scale of the RG flow.    
The coefficients $\Delta \mu_{d-k}$ have dimension $d-k$ and have the interpretation of a finite renormalization of the coefficient of $r^{d-k}$ between the UV and IR fixed points. The last term gives the change in the universal part of the EE: $\Delta A=A_{IR}-A_{UV}$, with $A$ the Euler trace anomaly coefficient for even dimensions, and $\Delta F=F_{IR}-F_{UV}$, with $F$ the constant term of the free energy of a $d$-dimensional Euclidean sphere. 

Concavity, Eq.~(\ref{sisi}), implies two relations between the short and long distance expansions for $\Delta S(a)$: 1) The slope of the $\Delta S(a)$ curve is bigger at the UV than at the IR; 2) Given that $\Delta S(0)=0$, the height at the origin of the tangent line at the IR has to be positive. 

The first requirement, comparing (\ref{oyo1}) and (\ref{even1}), and provided $\Delta < (d+2)/2$, gives place to the ``area theorem", that is, the decrease along the RG of the coefficient of the area term,\footnote{If 
$\Delta > (d+2)/2$ the area term at the $UV$ can be considered infinite because the slope of (\ref{oyo1}) diverges as $r\rightarrow 0$.}  
\be
\Delta \mu_{d-2}\le 0\,.\label{dfghj}
\ee
In $d=2$ the area coefficient is dimensionless and (\ref{dfghj}) coincides with the $c$-theorem. The area theorem was obtained in~\cite{Casini:2016udt} using monotonicity of the relative entropy.

The second requirement gives for $d=3$ the $F$-theorem,
\be
\Delta F\le 0\,,
\ee
and for $d=4$ the $a$-theorem,
\be
\Delta A\le 0\,.\label{mia}
\ee
For higher dimensions $d>4$ it gives
\be
\Delta \mu_{d-4}\ge 0\,.\label{tuya}
\ee 
The inequality does not constraint the sign of the subleading terms, in particular the universal terms, for $d>4$.  

In addition to these constraints that come from comparison of the UV and IR expansions, we have to check (\ref{sisi}) at the UV and infrared expansions themselves. At the IR we get again (\ref{mia}) and (\ref{tuya}) for $d\ge 4$. For $d=3$ we get information on the sign of the first subleading correction to the constant
\be
\Delta S^{d=3}_{IR}= \Delta \mu_1 r -\Delta F - \frac{k}{r^{\alpha}}+\ldots\,,
\ee
where the last term is purely infrared in origin and $\alpha$ is related to the leading irrelevant dimension of the operator driving the theory to the IR \cite{Liu:2012eea}. We get $k>0$ from (\ref{sisi}). This coincides with holographic calculations \cite{Liu:2013una}, and free field theory calculations \cite{Huerta:2011qi}. 
 At the UV we get that the sign of the coefficient $c_0$ in (\ref{oyo1}) is the same as the one of $\Delta-(d+2)/2$. This also agrees with holographic calculations \cite{Liu:2012eea}.

Notice that while the inequality (\ref{sisi}) saturates at the UV, it does not saturate at the IR for $d\ge 4$. The SSA inequality always saturates at the IR for regions smooth enough (with IR size curvatures) but this does not allow us to derive (\ref{sisi}) precisely because we are not allowed to convert wiggly spheres into spheres for these large wiggles.  

\section{Final remarks} 
\label{sec:concl}

We have found that the Markov property for EE on the plane, and on the light-cone for CFT's, has an origin that is essentially geometric. Because of that, this property extends to other quantities, e.g. the Renyi entropies; it does not depend on other specific properties that the EE has -- and the Renyi entropies generally do not have -- such as the SSA inequality. The Markov property together with Lorentz invariance determine the general form of the entropies on the light-cone for a CFT, and turns out to be related to dilaton effective actions in two less dimensions. The universal part is completely fixed by the coefficient $A$ of the conformal anomaly in even dimensions and is given by the Wess-Zumino anomaly action. For odd dimensions the universal part is just a constant $F$ for any region in the light-cone. 

Beyond cases that are conformal transformations of the null plane in Minkowski space for CFT's, we expect that the Markov property also holds for any QFT on an space-time having a bifurcate Killing horizon, and where the state is invariant under the Killing symmetry. This is because the Killing symmetry will squash all regions to the bifurcation and keep a covariant cutoff invariant, leading to constant entropies on the horizon. This includes for example, arbitrary QFT in de Sitter space for the de Sitter invariant state and regions on the cosmological horizon, and for regions on the horizon of stationary black holes for the Hartle-Hawking state.

The Markov property for the Renyi entropies extends the constraints on the density matrix beyond Markovianity. For finite systems, the Markov property for all Renyi entropies in subsystems $A$, $B$, $C$, 
\be
S_n(AB)+S_n(BC)=S_n(B)+S_n(ABC)\,, 
\ee  
can only be possible if the global state is of the form, $\rho_{ABC}=\rho_{AB_1}\otimes \rho_{B_2C}$, with $B_1$ and $B_2$ two subsystems partitioning $B$.
Hence, $\rho_{AC}=\rho_A\otimes \rho_C$ is a product. This suggests that the vacuum state is roughly a product over different null pencils in vacuum QFT, though this is not quite correct mathematically for a theory in $d>2$ and an interacting UV fixed point.  In this case, the algebras corresponding to finite regions on the light-cone (that do not generate a domain of dependence containing spacetime volume) actually have no degrees of freedom. Anyway, in the cases where this identification makes sense, free theories and CFTs in $d=2$, one can check that the structure of the vacuum is in fact a product state, rather than a more general Markovian state where classical correlations are allowed between $A$ and $C$. For free theories this is described in \cite{Wall:2011hj}, while for a CFT in $d=2$ the vacuum is a product across the two null directions. 

The present investigation started in the course of attempting to generalize the entropic proofs of the $c$ and $F$ theorems to $d=4$. In this sense it is intriguing that we have found that the entropies on the null cone are classified by dilaton effective actions, which are fundamental in the proof by Komargodski and Schwimmer of the $a$-theorem \cite{Komargodski:2011vj}. However, in the present case, the dilaton lives in $d-2$ dimensions rather than $d$ dimensions. This connection was also noticed by Solodukhin in \cite{Solodukhin:2013yha}. Another difference is that our non dynamical dilaton does not necessarily obey unitarity constraints. It would be interesting to investigate if this connection could be the base for extending the irreversibility theorems to dimensions higher than $d=4$.  

We have checked that the general expressions for the entropy on the cone hold holographically. It is surprising that exact holographic expressions can be found for the entropy of such a large class of regions, though we can understand the origin of this simplification from more general principles. We have discussed how this simplification also permeates to $\lambda^{-1}$ and $N^{-1}$ corrections. Holographically, the origin of all the simplifications is the fact that the entangling surface lies on a maximally symmetric null cone in the bulk. 

It would be interesting to obtain the expected form of the Renyi entropies on the cone from a direct calculation of the holographic Renyi entropies.  In this case we would have to deal with a (in principle) complicated Schwinger-Keldysh representation with Lorentzian conical defects in the bulk \cite{Dong:2016hjy} because we cannot use the Euclidean representation \cite{Dong:2016fnf,Lewkowycz:2013nqa} for generic regions living on the null cone. Our best guess is that the bulk manifold should still be locally AdS, in such a way as to allow to locate the defects on a fixed bulk null cone. If this is the case, the Markov property and the expected expansion of the Renyi entropies would hold by the same reasons discussed in this paper for the entropy.  

\section*{Acknowledgments}
We thank Xi Dong, Aitor Lewkowycz, Juan Maldacena, and Mark Van Raamsdonk for discussions.  We would like to dedicate this work to the memory of Joe Polchinski.
This work was partially supported by CONICET (PIP grant 11220150100299), CNEA, and Universidad Nacional de Cuyo, Argentina. H.C. acknowledges an ``It From Qubit" grant of the Simons Foundation. G.T. is also supported by ANPCYT PICT grant 2015-1224.

\appendix

\section{Lorentz invariant regularization using mutual information}
\label{Appendix:mutual}

In this Appendix we review the Lorentz invariant regularization of EE provided by the mutual information for any QFT in any dimension. This is discussed in detail for $d=3$ in \cite{Casini:2015woa}. We are restricting attention to smooth entangling surfaces, which is all we need in this paper. 

Consider a smooth entangling surface $\gamma$. We take a spatial unit vector $\eta$ normal to $\gamma$, and a function $\epsilon(x)$ on $\gamma$, which is a smoothly varying short distance on the surface. We will later take the limit $\epsilon(x)\rightarrow 0$, and impose that in this limit the derivatives of $\epsilon(x)$ approach zero at the same rate as $\epsilon(x)$. We can construct two spatial surfaces, one on each side of $\gamma$, by using the elements of the ``framing" $(\eta,\epsilon)$,
\bea
\gamma^+&=& \gamma+\frac{\epsilon}{2} \eta\,,\\ 
\gamma^-&=& \gamma-\frac{\epsilon}{2} \eta\,.
\eea   

The idea is to use the mutual information $I(\gamma^+,\gamma^-)$ as a regularization of the entropy. More precisely we take
\be
S_{\textrm{reg}}(\gamma,\eta,\epsilon)=\frac{I(\gamma^+,\gamma^-)}{2}=\frac{1}{2}\left(S(\gamma^+)+S(\gamma^-)-S(\gamma^+\cup \gamma^-)\right)\,.\label{ffkl}
\ee
For the Renyi entropies we use analogously the mutual Renyi entropies $I_n(\gamma^+,\gamma^-)=S_n(\gamma^+)+S_n(\gamma^-)-S_n(\gamma^+\cup \gamma^-)$. The $1/2$ factor in (\ref{ffkl}) takes into account that the mutual information for complementary regions in a global pure state is twice the entropy. An important point is that the mutual information is regularization independent, that is, taking the continuum limit of any regularization for the entropies on the right hand side of (\ref{ffkl}) should give the same finite result. Hence, $S_{\textrm{reg}}$ is a quantity that belongs to the continuum theory, and in particular is Lorentz invariant in vacuum. The particular symmetric framing on both sides of $\gamma$ in (\ref{ffkl}) gives the same regularized entropy for complementary regions, as expected property for the entropy of global pure states. 

However, $S_{\textrm{reg}}$ depends on the framing, that includes the vector field $\eta$, and it is not a function of the entangling surface $\gamma$ alone. In order to get rid of this unwanted framing dependence we note that as we are taking the $\epsilon\rightarrow 0$ limit, we only retain non-positive powers of $\epsilon$. The dependence on $\eta$ can only show up in the divergent terms. As these are produced by ultralocal entanglement between regions arbitrarily close to both sides of $\gamma$, these contributions can be written as integrals of local geometrical terms along $\gamma$. Now we can just subtract these terms to eliminate the frame dependence
\be
S_{\textrm{reg}}(\gamma)=S_{\textrm{reg}}(\gamma,\eta,\epsilon)-\textrm{local divergent terms}\,.
\label{misia}\ee 
 This is finite, Lorentz invariant, and completely defined by the theory itself. It can be thought of  as a ``minimally  subtracted" entropy.

\begin{figure}[t]
\begin{center}  
\includegraphics[width=0.65\textwidth]{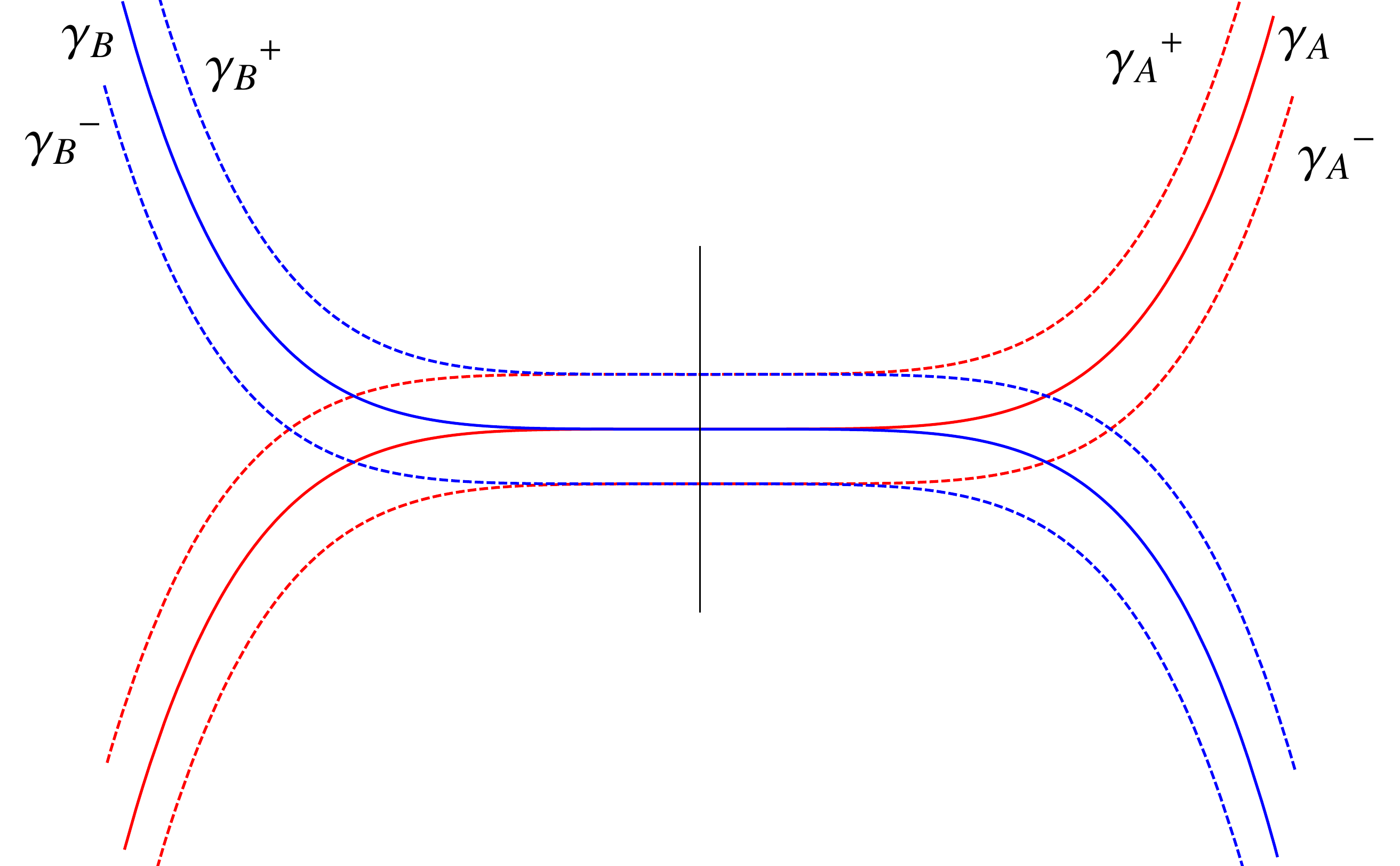}
\captionsetup{width=0.9\textwidth}
\caption{Strong subadditivity of the regularized entropies of two surfaces $\gamma_A$ and $\gamma_B$ with smooth intersection and union. The framings of $\gamma_A$ and $\gamma_B$ can be chosen such that they are compatible, i. e., they can be split along the black line in the middle, and reconnect to form the framings of $\gamma_{A\cap B}$ and $\gamma_{A \cup B}$.}
\label{fig:mutuals}
\end{center}  
\end{figure}

 While $S_{\textrm{reg}}(\gamma)$ does not have the property of being positive for arbitrary regions,  it does retain some other important properties of entropy. The symmetry between complementary regions is one of these properties, and the other is strong subadditivity. This is shown as follows.
 
 First we take two regions $\gamma_A$ and $\gamma_B$ with smooth intersection $\gamma_A\cap \gamma_B$ and  union $\gamma_A\cup \gamma_B$. Then we take compatible framings, as in Fig.~\ref{fig:mutuals}. We expect that the thin strip terms exactly cancel in
 \be
 S(\gamma_A^+\cup \gamma_A^-)+S(\gamma_B^+\cup \gamma_B^-)-S(\gamma_{A\cap B}^+\cup \gamma_{A\cap B}^-)-S(\gamma_{A\cup B}^+\cup \gamma_{A\cup B}^-)=0\,. \label{cancel}
 \ee     
This is because these strip entropies should be taken as expansions in inverse powers of $\epsilon$, and these expansions should be local and extensive along the strips. Thinking in terms of the Renyi entropies, this should be a property of the operator product expansion of surface twist operators. The cancellation (\ref{cancel}) gives place to the strong subaditivity of the regularized entropies just because the entropies themselves are strong subadditive,
\bea 
\nonumber && S_{\textrm{reg}}(\gamma_A)+S_{\textrm{reg}}(\gamma_B)-S_{\textrm{reg}}(\gamma_{A\cap B})-S_{\textrm{reg}}(\gamma_{A\cup B})=\frac{1}{2}\left(S(\gamma^+_A)+S(\gamma_B^+)-S(\gamma^+_{A\cap B})-S(\gamma_{A\cup B}^+)\right.\\&& \left. +S(\gamma^-_A)+S(\gamma_B^-)-S(\gamma^-_{A\cap B})-S(\gamma_{A\cup B}^-)\right)\ge 0\,.
\eea
In a sense, since the entropies are strong subadditive, subtracting the frame dependent terms cannot change this fact because divergent terms are always Markovian for smooth enough surfaces. For holographic theories $S_{\textrm{reg}}$ is just the entropy with the usual Lorentz invariant cutoff and the divergent terms subtracted.

\section{Extrinsic curvatures on the null cone}
\label{Appendix:uno}

In this Appendix we argue that the extrinsic curvatures on the null cone do not give rise to additional geometric invariants besides those studied in Sec.~\ref{sec:qft}.

We have a surface $r=\gamma(\Omega)$ on the null cone $r^{+}=0$. 
 Let us define $n^{(1)}=\hat{r}-\hat{t}$ as the null vector parallel (and orthogonal) to the cone. Let $q=1/2(\hat{r}+\hat{t})$, with $q^2=0$, $q\cdot n^{(1)}=1$.  The orthogonal vector space to $\gamma$ is formed by $n^{(1)}$ and another null vector $n^{(2)}$ given by
\be
n^{(2)}=q-\frac{1}{2} (\nabla \gamma)^2 n^{(1)}- \nabla \gamma\,.
\ee
This is normalized such that $n^{(1)}\cdot n^{(2)}=1$. 

The extrinsic curvatures  corresponding to $n^{(i)}$ are defined by 
\be
K^{(i)}_{\mu\nu}=P^\alpha_\mu P^\beta_\nu \nabla_\alpha n^{(i)}_\beta\,, 
\ee
with 
\be
P^\alpha_\beta=g^\alpha_\beta-n^{(1)\alpha} n^{(2)}_\beta-n^{(2)\,\alpha} n^{(1)}_\beta
\ee
the projector onto the tangent space to $\gamma$.

The vector $n^{(1)}_\mu=x_\mu/(|\vec{x}|)$ in Cartesian coordinates, and we get  
\be
K^{(1)}_{\mu\nu}=\frac{\, g^{\textrm{int}}_{\mu\nu}}{\gamma}\,,\label{sese}
\ee
with $g^{\textrm{int}}$ the intrinsic metric on $\gamma$. 
The other extrinsic curvature is
\be
K^{(2)}_{\mu\nu}=\frac{1}{2}\frac{\, g^{\textrm{int}}_{\mu\nu}}{\gamma}-\frac{1}{2} (\nabla \gamma)^2 \frac{\, g^{\textrm{int}}_{\mu\nu}}{\gamma}
-(\nabla_\mu\nabla_\nu \gamma)^{\textrm{int}}  \,,
\ee
where we have used that the derivatives of $\hat{t}$ are zero and hence the gradient of $q$ is one half  that of $n^{(1)}$. In the last term the second derivatives are finally projected onto the parallel subspace. We have that $\nabla^{\textrm{int}}_\mu \gamma=\nabla \gamma + (\nabla \gamma)^2 n^{(1)}$ because this vector is parallel to the surface. Hence  $(\nabla_\mu\nabla_\nu \gamma)^{\textrm{int}}=\nabla^{\textrm{int}}_\mu\nabla^{\textrm{int}}_\nu \gamma- (\nabla \gamma)^2 g^{\textrm{int}}_{\mu\nu}/\gamma $. 
Using angular coordinates for the surface we have the intrinsic metric $ds^2=\gamma(\Omega)^2 d\Omega^2$. We have, writing all covariant derivatives and contractions with respect to the metric $g_{\mu\nu}$ of the unit sphere,
\be
K^{(2)}_{\mu\nu}=\frac{1}{2} \gamma g_{\mu\nu}-\frac{1}{2} (\nabla \gamma)^2 \frac{\, g_{\mu\nu}}{\gamma}-\nabla_\mu\nabla_\nu \gamma + 2 \frac{\nabla_\mu \gamma \nabla_\nu \gamma}{\gamma} \,.
\ee

On the other hand, using formulae for the conformal transformations, the Ricci tensor and Ricci scalar are given by
\bea
R^{\textrm{int}}_{\mu\nu}&=&R_{\mu\nu}-(d_\perp-2)\frac{\nabla_\mu\nabla_\nu\gamma}{\gamma}+2(d_\perp-2)\frac{\nabla_\mu \gamma\nabla_\nu \gamma}{\gamma^2}+g_{\mu\nu}\left((3-d_\perp)\frac{(\nabla \gamma)^2}{\gamma^2}-\frac{\nabla^2 \gamma}{\gamma}\right)    \nonumber\,,\\
g_{\mu\nu}^{\textrm{int}} R^{\textrm{int}}&=&g_{\mu\nu}\left(R-2(d_\perp-1)\frac{\nabla^2 \gamma}{\gamma}+(d_\perp-1)(4-d_\perp)\frac{(\nabla \gamma)^2}{\gamma^2} \right)  \,.
\eea
Using that for the unit sphere $R_{\mu\nu}=(d_\perp-1)g_{\mu\nu}$ and $R=d_\perp(d_\perp-1)$ we have
\be
K^{(2)}_{\mu\nu}=\frac{\gamma}{d_\perp-2}\left(R_{\mu\nu}^{\textrm{int}}-\frac{1}{2(d_\perp-1) }g_{\mu\nu}^{\textrm{int}}R^{\textrm{int}}\right)\,.\label{sasa}
\ee

Therefore, from (\ref{sese}) and (\ref{sasa}) we conclude that using the extrinsic curvatures of $\gamma$ we can not form additional invariants to the ones formed with the intrinsic geometry of $\gamma$ on the null cone. For example, the invariant multiplying the type $B$ anomaly coefficient in Solodukhin's formula \cite{Solodukhin:2008dh} for the universal logarithmic term of the entanglement entropy in $d=4$ vanishes,  
\be
K^{(1)}_{\mu\nu}K^{(2)\,\mu\nu}-\frac{1}{2}K^{(1)\,\mu}_\mu K^{(2)\,\mu}_\mu=0\,.
\ee
Hence only the $A$ anomaly contributes on the cone. 

\section{EE for wiggly spheres in holographic RG flows}\label{app:RG}


We are going to compute holographically terms in the entropy induced by a mass parameter in the difference between the entropy of a sphere of radius $R$ and a wiggly sphere centered around the same radius. We work in $d=4$ for concreteness. As a model for wiggly sphere we consider
\be
\gamma^{-1}=R^{-1} \left(1+ \frac{a}{\sqrt{2}} \, (Y_{lm}(\Omega)+Y_{lm}^*(\Omega))\right)\,.  
\ee
We are looking for the limit of small wiggle size, $l\rightarrow \infty$, $a\rightarrow 0$, and, as in the proof of the $a$-theorem, we take the size of the wiggles of the order of their width, $a\sim l^{-1}$. The result is independent of $m$. We choose $m=0$. 

The solution for the extremal surface for the UV CFT is given by (\ref{eq:solrt})
\be\label{eqqqb}
(\t r(\theta, \Omega))^{-1} =  R^{-1}\left(1+a\,Y_{l0}(\Omega)\,\, \frac{\sqrt{\pi} \Gamma(3+l)}{2^{2+l}\Gamma(\frac{3+2l}{2})}\,\,(\cos \theta)^l\,{}_2F_1(\frac{l-1}{2}, \frac{l}{2}, \frac{3}{2}+l,\cos^2\theta)\right)\,.
\ee
The function of $l$ and $\theta$ multiplying $a Y_{l0}$ has value $1$ for $\theta=0$ and decays exponentially fast with $l$ large for fixed $\theta >0$. It is not exponentially suppressed only for $\theta\lesssim l^{-1}$. This means that the deformation due to the wiggles on the minimal surface decays exponentially fast towards the interior of AdS, and, for small wiggle width, are only relevant  near the AdS boundary. This means their contribution is dominated (except for terms exponentially small in the inverse wiggle size) by the UV fixed point. Hence, in an holographic calculation we can just use the UV perturbed AdS metric to compute the effect of the mass deformation on the wiggles. 

Near the boundary the metric is deformed to leading order as
\be
ds^2= \frac{dx^2 +dz^2 (1- g^{2} z^{2 \alpha})}{z^2} \,,
\ee 
where $g$ is proportional to the coupling constant and $\alpha=d-\Delta$, with $\Delta$ the scaling dimension of the operator producing the RG flow.  In terms of the $\tilde{r}, \theta$ coordinates, the change in the metric is
\be
\delta ds^2=- g^{2} (\tilde{r} \sin(\theta))^{2 \alpha-2} \left(\frac{d\tilde{r}^-}{2}\sin(\theta)+\frac{d\tilde{r}^+}{2}\sin(\theta)+\tilde{r} \cos(\theta) d\theta\right)^2\,.
\ee
The variation of the area due to the variation of the metric is
\be
\delta \mc A=\frac{1}{2}\int d\Omega\,d\theta\, \sqrt{h}\, g^{\mu\nu} \delta h_{\mu\nu}\,,
\ee
where $h_{\mu\nu}$ is the induced metric on the surface, and the computation is over the unperturbed surface. 

Then we get for the difference of entropies between wiggly and normal spheres, to leading order in $g^2$,
\be
\Delta \mc A=\delta \mc A_{\textrm{wiggly}}-\delta \mc A_{\textrm{sphere}}=-\frac{g^{2}}{2}\int d\Omega\,d\theta\, \cos(\theta)^4\sin(\theta)^{2 \alpha-3}\,\Delta(\tilde{r} )^{2 \alpha}  \,.  
\ee
The factor $\Delta(\tilde{r} )^{2 \alpha}$ decays exponentially towards the bulk and makes the perturbative expansion on the metric deformation valid.   Using (\ref{eqqqb}), and expanding for small wiggly size to second order to get a non trivial angular integral, we get
\be
\Delta \mc A=\alpha\,(\alpha-1)\,a^2\,g^{2}\, R^{2\alpha}\int_0^{\pi/2} d\theta\, \cos(\theta)^{4+2 l}\sin(\theta)^{2 \alpha-3}\,\left( \frac{\sqrt{\pi} \Gamma(3+l)}{2^{2+l}\Gamma(\frac{3+2l}{2})}\,_2F_1(\frac{l-1}{2}, \frac{l}{2}, \frac{3}{2}+l,\cos^2\theta)\right)^2   \,.  
\ee
The integrand is proportional to $\theta^{2 \alpha-3}$ for small $\theta$. Then the integral diverges for $\Delta\ge 3$, which is the onset of massive divergent area terms in $d=4$. The divergences give place to local terms that are Markovian and can be subtracted. For $\Delta< 3$, we get a finite integral with the following behavior for large $l$
\be\label{eq:dA}
\Delta \mc A\sim\,a^2 \,l^{-2 (3-\Delta)}\,\,g^{2}\, R^{2\alpha} \, \hspace{1cm} \Delta<3\,.  
\ee
This clearly vanishes in the limit of small wiggle size and width.
For $4>\Delta>3$ we have, once the divergence for $\theta\rightarrow 0$ has been subtracted,   the same result (\ref{eq:dA}).
Since we are taking the limit of small wiggles with fixed slope, $a\sim l^{-1}$, this term also vanishes in the limit of small wiggles. 
These terms represent the change of the non local term (\ref{oyo1}) due to the wiggles.

\bibliography{EE}{}
\bibliographystyle{utphys}

\end{document}